# QUANTUM FIELD THEORETIC DERIVATION OF THE EINSTEIN WEAK EQUIVALENCE PRINCIPLE USING EMQG THEORY

**Tom Ostoma and Mike Trushyk**

48 O'HARA PLACE, Brampton, Ontario, L6Y 3R8
emqg@rogerswave.ca

Thursday, February 25, 1999

**ACKNOWLEDGMENTS**

We wish to thank L. Walker for his help in proof reading this document.

# ABSTRACT

*We provide a quantum field theoretic derivation of Einstein's Weak Equivalence Principle (WEP) of general relativity using a new quantum gravity theory proposed by the authors called Electro-Magnetic Quantum Gravity or EMQG (ref. 1). EMQG is manifestly compatible with Cellular Automata (CA) theory (ref. 2 and 4), and is also based on a new theory of inertia (ref. 5) proposed by R. Haisch, A. Rueda, and H. Puthoff (which we modified and called Quantum Inertia, QI). QI states that classical Newtonian Inertia is a property of matter due to the strictly local electromagnetic force interactions of <u>each</u> of the (electrically charged) elementary particles of the mass (masseon particles) with the surrounding (electrically charged) virtual particles (virtual masseons) of the quantum vacuum. The sum of all the tiny electromagnetic forces (photon exchanges with the vacuum) originating in each charged elementary particle of the accelerated mass is the source of the <u>total inertial force</u> of a mass which opposes accelerated motion in Newton's law 'F = MA'. Therefore, classical Newtonian inertia follows from the basic principles of quantum field theory. The paradoxes that arise in acceleration (Mach's principle) are resolved, and Newton's laws of motion are now understood within quantum field theory.*

*We invoked Einstein's principle of equivalence of inertial and gravitational mass to understand the origin of gravitational mass from the perspective of quantum inertia. We found that gravity also involves the same 'inertial' electromagnetic force component that exists in inertial mass. We propose that Einstein's general relativistic Weak Equivalence Principle (WEP) originates from common 'lower level' quantum vacuum processes occurring in both gravitational mass and inertial mass in accordance with the principles of quantum field theory. Gravitational mass results from the interactions of <u>both</u> the electromagnetic force (photon exchanges) and the <u>pure</u> gravitational force (graviton exchanges) on matter, acting <u>simultaneously</u>. However, inertial mass is strictly the result of the electromagnetic force process only, as given by quantum inertia principle (with negligible graviton processes present). Under gravitation, a test mass near the earth exchanges gravitons with the earth. However, the surrounding (electrically charged) virtual particles <u>also</u> exchange gravitons with the earth, causing the virtual particles to accelerate (fall) towards the earth. A test mass does moves under the influence of the direct graviton exchanges, but more importantly also under the influence of the falling (electrically charged) virtual particles of the quantum vacuum, which dominates in the total force exchange process with electromagnetic interactions. Thus, a test mass under gravity 'sees' that the quantum vacuum accelerate in the same way as it did when the test mass was subjected to acceleration alone. Thus, equivalence arises from the reversal of the acceleration vectors of the quantum vacuum with respect to a test mass undergoing acceleration as compared to a test mass subjected to a gravitational field. In accelerated frames, it is <u>the mass that accelerates</u>; inside gravitational fields it is <u>the virtual particles of the quantum vacuum that accelerates</u>. However, the direct graviton exchanges with the test mass and the earth upsets perfect equivalence, with gravitational mass of an object being slightly larger than inertial mass. This violation of the WEP (with three other experimental tests of EMQG) might be detectable experimentally.*

*All elementary (fermion) particles consist of combinations of just <u>one</u> fundamental matter (and anti-matter) particle called the 'masseon' particle. The masseon has one, fixed (smallest) quanta of mass (similar to the idea of a quanta of electric charge), which we call low level 'mass charge'. The masseon also carries either a positive or negative (smallest) quanta of electric charge. The masseon particle generates a fixed flux of gravitons, with a flux rate being <u>completely unaffected</u> by relativistic motion. The graviton is the vector boson exchange particle of the <u>pure</u> gravitational force interaction. In EMQG, the physics of graviton exchanges is nearly identical to the photon exchange process in QED, with the same concept of positive and negative gravitational 'mass charge' carried by masseons and anti-masseons respectively (this is a gross violation of the equivalence principle for anti-matter). The ratio of the graviton to photon exchange force coupling is about $10^{-40}$. In QED, the quantum vacuum consists of virtual electrons, virtual anti-electrons, and virtual photons. In EMQG, the quantum vacuum consists of virtual masseons, virtual anti-masseons, and virtual gravitons, which also posses both positive and negative electrical charge and positive and negative 'mass charge'. There are almost*

*equal numbers of virtual masseon and anti-masseon particles existing in the quantum vacuum everywhere, and at any given time. This is why the cosmological constant is very close to zero in the universe; there is an equal proportion of attractive and repulsive gravitational forces in the quantum vacuum.*

# **TABLE OF CONTENTS**

# 1. INTRODUCTION

*"I have never been able to understand this principle (principle of equivalence) ... I suggest that it be now buried with appropriate honors."*

**- Synge:  Relativity- The General Theory**

The principle of equivalence is one of the founding postulates of general relativity theory. When stated in it's weaker form, objects of different mass fall at the same rate of acceleration in a uniform gravity field. Alternatively, it means that the inertial mass (mass defined by Newton's law of motion: $m_i = F_i/g$) is ***exactly*** equal to the gravitational mass (mass defined by Newton's universal gravitational law $m_g = (F_g \, r^2) / (GM)$ ). The equivalence principle requires that $m_i = m_g$. It is still not understood why inertial mass exists in the first place, or in other words, why a mass opposes acceleration with an inertial force. More importantly, it is also not known why these two different physical definitions for inertial and gravitational mass exist (instead of only one definition), and why they give the *same* numerical value for a mass of any material composition and energy content.

Imagine that you are standing on the surface of the earth. Gravity appears like a static force holding your mass to the surface. Yet, when you are standing in an accelerated rocket moving with an acceleration of 1 g, the principle of equivalence states that there is an identical force exerted against the rocket floor by your inertial mass which is now caused by your dynamic accelerated motion. Why should there be such a deep connection between what appears to be two completely different physical phenomena: a static force and a dynamic force?

Lacking any deeper understanding of this question, some physicists prefer to accept this law as the fundamental way in which the universe operates. Other physicists maintain that the origin of the principle of equivalence is one of the deepest, unsolved mystery of modern physics. This paper provides a quantum field theoretic solution to the problem of the origin of the principle of equivalence. In other words, we show how the principle of equivalence turns out to be a quantum process that results solely from the activities of quantum particles interacting with quantum particles, while obeying the general laws of quantum field theory. It is also an invitation to explore a new theory of gravity, called ElectroMagnetic Quantum Gravity, or EMQG. EMQG is based on a new understanding of both inertia and the principle of equivalence that exists on the quantum particle distance scales (ref. 1).

How is the principle of equivalence defined? There are two main formulations of the principle of equivalence: the Weak Equivalence Principle (WEP), and the Strong Equivalence Principle (SEP). The strong equivalence principle states that the results of *any given physical experiment* will be precisely identical for an accelerated observer in free space as it is for a non-accelerated observer in a perfectly uniform gravitational field. A weaker form of equivalence principle restricts itself to the laws of motion of masses only. In other words, the weak principle of equivalence states that only the laws of motion of a mass on the earth is identical to the laws of motion of the same mass inside an accelerated rocket (at 1g). Technically, when comparing the equivalence of a mass in a rocket to a mass on the earth, we assume that the motion is restricted to short distance (mathematically, at a point) on the earth, where gravity does not vary with height. The WEP implies that objects of different mass, different material composition, and different energy content fall at the same rate of acceleration on the earth (as they do in a rocket accelerating at 1g). How do we know that the WEP is true?

One of the earliest experimental tests of equivalence was the historical experiment by Galileo in Pisa, where two objects of significantly different mass were dropped off the leaning tower of Pisa, and observed to land on the ground roughly at the same time. The equivalence principle implies that these two different masses should fall at exactly the same rate (as they do inside an accelerated rocket). Since then, the equivalence of inertia and gravitational mass has been verified to a phenomenal accuracy of one part in about $10^{-15}$ (ref 24). Einstein is credited with the elevation of the equivalence principle to a fundamental symmetry of nature in 1915.

Conventional wisdom in physics assumes that the strong principle of equivalence is *exact*, and somehow reflects a fundamental aspect of nature. It is assumed to be applicable under any physical circumstance. It is believed to hold true at the elementary particle level, and under enormously large gravitational fields such as on a neutron star. As a consequence, Einstein's general relativity theory (which is based on this principle) is also assumed to hold true under any physical condition. The principle of equivalence has been tested under a wide variety of gravitational field strengths and distance scales. It has been tested with different material types (ref. 6). It has been tested to an extremely high precision for laboratory bodies (up to 3 parts in $10^{-12}$). It has been tested to 1 part in $10^{-12}$ for the acceleration of the moon and earth towards the sun. It has been tested for elementary particles, such as the neutron.

Yet, after 80 years of close scrutiny the principle of equivalence has still remained only a postulate of general relativity. It cannot be proven from more fundamental principles. Some of the better literature on general relativity have drawn attention to this fact, and admit that no explanation can be found as to; "why our universe has a deep and mysterious connection between acceleration and gravity" (ref. 8). We will show that the principle of equivalence is in fact just an approximation, albeit an extremely close approximation. We will also show that there is actually a tiny imbalance in equivalence, with the gravitational mass of an object being (every so slightly) larger than the inertial mass. This effect is magnified when comparing the free fall times of very large mass to that of an extremely tiny mass. Furthermore, this effect may be measured experimentally in the near future. We will also show that in some extremely rare physical circumstances the equivalence principle does not hold at all!

In order to understand the equivalence principle (and inertia), we must have an understanding of the basic concepts of Electromagnetic Quantum Gravity (EMQG). The motivation for the development of EMQG was the consideration that the universe may be a Cellular Automata. In the process of developing EMQG theory with this goal in mind, we discovered the hidden processes that are responsible for the principle of equivalence. Therefore, before we can present our derivation of the equivalence principle, we must first very briefly review EMQG theory. Refer to Reference 1 for the complete work.

## 2. INTRODUCTION TO ELECTROMAGNETIC QUANTUM GRAVITY

*"The interpretation of geometry advocated here cannot be directly applied to submolecular spaces … it might turn out that such an extrapolation is just as incorrect as an extension of the concept of temperature to particles of a solid of molecular dimensions"*
                                                                                                             *A. Einstein (1921)*

Various attempts at the unification of general relativity with quantum theory have not been entirely successful in the past, because these theories do not grasp the true nature of inertia and the hidden quantum physical processes behind Einstein's principle of equivalence. In developing a theory of quantum gravity, one might ask which of the existing approaches to quantum gravity is more relevant or fundamental, quantum field theory or classical general relativity (with conventional 4D space-time continuum)? Currently it seems that both theories are generally not compatible with each other.

Based on a postulate that the universe operates like a Cellular Automata (CA) (ref. 2 and 4), we have taken the position that quantum field theory is actually in closer touch to the workings of our universe. General relativity is taken as a global, classical description of space-time, gravity and matter. General relativity reveals the large-scale patterns and organizing principles that arise from the hidden quantum processes existing on the quantum distance scales. Quantum field theory is closer to revealing the inner fundamental workings of the universe, and tells us that all forces originate from a quantum particle exchange process. These particle exchanges transfer momentum from one quantum particle to another. The huge numbers of particles exchanged produce a smooth force interaction. The exchange process is universal, and applies to electromagnetic, weak and strong nuclear forces, and also for gravitational force. The generic name given to the force exchange particle is the 'vector boson' particle.

We have developed a quantum theory of gravity called ElectroMagnetic Quantum Gravity (or EMQG, reference 1) that is manifestly compatible with the Cellular Automata model of the universe. EMQG supports the view that gravity is based on particle exchange processes in accordance with the general principles of quantum field theory. EMQG is based on the photon (the exchange particle for electromagnetic force) and the graviton (the exchange particle for the *pure* gravitational force). What is unique about EMQG theory is that gravitation involves <u>both</u> the photon and graviton exchange particles operating at the same time, where now the photon plays a very important role in gravity! In fact, the photon exchange process dominates all gravitational interactions and is, in most part, responsible for the principle of equivalence. The photon particle is also responsible for another property that all matter possesses, the Newtonian inertia.

In order to formulate a theory of quantum gravity, a mechanism must be found that produces the gravitational force, while somehow being linked to the principle of equivalence. In addition, this mechanism should naturally lead to 4D space-time curvature and should be compatible with the principles of general relativity theory. Nature has another long-range force called electromagnetism, which has been described successfully by the principles of quantum field theory. This well-known theory is called Quantum ElectroDynamics (QED), and this theory has been tested for electromagnetic phenomena to an unprecedented accuracy. It is therefore reasonable to assume that gravitational force should be a similar process, since gravitation is also a long-range force like electromagnetism. However, a few obstacles lie in the way, which complicate this line of reasoning.

First, gravitational force is observed to be <u>always</u> attractive! In QED, electrical forces are attractive and repulsive. As a result of this, there are an equal number of positive and negative charged virtual particles in the quantum vacuum (section 7.1) at any given time because virtual particles are always created in equal and opposite charged particle pairs. Thus, there is a balance of attractive and repulsive electrical forces in the quantum vacuum, and the quantum vacuum is electrically neutral, overall. If this were not the case, the electrically charged virtual charged particles of one charge type in the vacuum would dominate over all other physical interactions involving real matter, due to the enormous number of vacuum particles involved.

Secondly, QED is formulated in a relativistic, flat 4D space-time with no curvature. In QED, electrical charge is defined as the fixed rate of emission of photons (strictly speaking, the fixed probability of emitting a photon) from a given charged particle. Electromagnetic forces are caused by the exchange of photons, which propagate between the charged particles. The photons transfer momentum from the source charge to the destination charge, and travel in flat 4D space-time (assuming no gravity). From these basic considerations, a successful theory of quantum gravity should have an exchange particle or graviton, which is emitted from a mass particle at a fixed rate as in QED. This is called 'mass charge', and is analogous to electrical charge in QED. Therefore, the graviton transfers momentum from one mass to another, which is the root cause of gravitational force. Yet, the graviton exchanges must somehow produce disturbances of the normally flat space and time, when originating from a very large mass.

Since mass is known to vary with velocity (special relativity), one might expect that 'mass charge' of a particle must also vary with velocity. In QED the electromagnetic force exchange process is governed by a fixed, universal constant ($\alpha$) which is not affected by anything like motion (more will be said about this point later). Should this not be true for graviton exchange process in quantum gravity as well? It is also strange that gravity, which is also a long-range force, is governed by a similar mathematical law as found in Coulomb's Electrical Force law. Coulomb's Electric Force law states: $F = KQ_1Q_2/R^2$, and Newton's Gravitational Force law: $F=GM_1M_2/R^2$. This certainly suggests that there is a deep connection between gravity and electromagnetism. Yet, gravity supposedly has no counterpart to negative electrical charge. Thus, this leads us to believe that there is no such thing as negative 'mass charge' for gravity. Furthermore, QED also has no analogous phenomena as the principle of equivalence. Why should gravity be connected with the principle of equivalence, and thus inertia, and yet no analogy of this phenomena exists for electromagnetic phenomena?

To answer the question of negative 'mass charge', EMQG postulates the existence of negative 'mass charge' for gravity, in close analogy to electromagnetism. Furthermore, we claim that this property of matter is possessed by all anti-particles that carry mass. Therefore anti-particles, which are opposite in electrical charge to ordinary particles, are also opposite in 'mass charge'. In fact, negative 'mass charge' is not only abundant in nature, it comprises nearly half of all the mass particles in the form of 'virtual' particles in the universe! The other half exists as positive 'mass charge', also in the form of virtual particles. Furthermore, all familiar ordinary (real) matter comprises only a tiny fraction of the total 'mass charge' in the universe, and is experimentally found to be almost all positive! Real anti-matter seems to be very scarce in nature, and no search for it in the cosmos has revealed any to date.

Both positive and negative 'mass charge' appear in huge numbers in the form of virtual particles, which quickly pop in and out of existence in the quantum vacuum (section 4), everywhere in empty space. We will see that the existence of negative 'mass charge' is the key to the solution to the famous problem of the cosmological constant, which is one of the great unresolved mysteries of modern physics. Finally, we propose that the negative energy, or the antimatter solution of the famous Dirac equation of quantum field theory is also the ***negative 'mass charge' solution***.

Previous attempts at quantizing the gravitational field have been made using the principles of quantum field theory. They focused on using the graviton force exchange particle, alone, as the quanta of the gravitational field, in direct analogy with the quantization of electromagnetic fields with photons. The graviton particle is chosen with the right mathematical characteristic to quantize gravity in accordance with quantum field theory and general relativity. These attempts however, fail to account for the origin of space-time curvature. Specifically, how does a graviton 'produce' curvature when propagating from one mass to another? Does the graviton move in an already existing 4D space-time curvature? If it does, how is the space-time produced by the graviton? If not, how is 4D space-time curvature produced? In other words, if the 4D space-time curvature is not caused by the graviton exchanges, then what is the cause?

Furthermore, why do the virtual particles of the quantum vacuum ***not*** contribute a nearly infinite amount of curvature to the whole universe? After all, the force of gravity is universally attractive. According to quantum field theory, virtual gravitons should exist in huge numbers in the quantum vacuum, and should therefore contribute huge amounts of attractive forces and a large amount of space-time curvature. This infamous question is known as the problem of the cosmological constant.

Does graviton exchange processes get affected by high velocity motion (with respect to some other reference frame)? In other words, do the number of gravitons exchanged increase as the velocity of the mass increases, as seems to be required by the special relativistic mass increase with velocity formula? Why does the state of motion of an observer near a gravitational field affect his local 4D space-time curvature? For example, why does an observer in free fall near the earth affect his local space-time conditions in such a way as to match an observer in far space (who lives in flat space-time)?

EMQG was developed to answer these questions, which have remained unresolved in existing quantum gravity theories. Before we can present our derivation of the equivalence principle, we present a brief review of the important results of EMQG theory. The full details are available in reference 1.

### 3. BRIEF REVIEW OF ELECTROMAGNETIC QUANTUM GRAVITY

*"All (the universe) is numbers"*

*- Pythagoras*

*"Subtle is the lord…"*

*- Einstein*

We have developed a new approach to the unification of quantum theory with general relativity referred to as ElectroMagnetic Quantum Gravity or EMQG (ref. 1). EMQG has its origins in Cellular Automata (CA) theory (ref. 2 and 4), and is also based on the new theory of inertia that has been proposed by R. Haisch, A. Rueda, and H. Puthoff (ref. 5) known here as the HRP Inertia theory. These authors suggested that inertia is due to the strictly local force interactions of charged matter particles with their immediate background virtual particles of the quantum vacuum. They found that inertia is caused by the magnetic component of the Lorentz force, which arises between what the author's call the charged 'parton' particles in an accelerated reference frame interacting with the background quantum vacuum virtual particles. The sum of all these tiny forces in this process is the source of the resistance force opposing accelerated motion in Newton's F=MA. We have found it necessary to make a small modification of HRP Inertia theory as a result of our investigation of the principle of equivalence. The modified version of HRP inertia is called "Quantum Inertia" (or QI). In EMQG, a new elementary particle is required to fully understand inertia, gravitation, and the principle of equivalence (described in the next section). This theory also resolves the long outstanding problems and paradoxes of accelerated motion introduced by Mach's principle, by suggesting that the vacuum particles themselves serve as Mach's universal reference frame (for <u>acceleration</u> only), without violating the principle of relativity of constant velocity motion. In other words, our universe offers no observable reference frame to gauge inertial frames, because the quantum vacuum offers no means to determine absolute constant velocity motion. However for accelerated motion, the quantum vacuum plays a very important role by offering a resistance to acceleration, which results in an inertial force opposing the acceleration of the mass. Thus, the very existence of inertial force reveals the absolute value of the acceleration with respect to the net statistical average acceleration of the virtual particles of the quantum vacuum. Reference 14 offers an excellent introduction to the motion of matter in the quantum vacuum, and on the history of the discovery of the virtual particles of the quantum vacuum.

There have been various clues to the importance of the state of the virtual particles of the quantum vacuum, with respect to the accelerated motion of real charged particles. One example is the so-called Davies-Unruh effect (ref. 15), where uniform and linearly accelerated charged particles in the vacuum are immersed in a heat bath, with a temperature proportional to acceleration (with the scale of the quantum heat effects being very low). However, the work of reference 5 is the first place we have clearly seen the identification of inertial forces as the direct consequence of the interactions of real matter particles with the virtual particles of the quantum vacuum.

It has even been suggested that the virtual particles of the quantum vacuum are involved in gravitational interactions. The prominent Russian physicist A. Sakharov proposed in 1968 (ref. 16) that Newtonian gravity could be interpreted as a van der Waals type of force induced by the electromagnetic fluctuations of the virtual particles of the quantum vacuum. Sakharov visualized ordinary neutral matter as a collection of electromagnetically, interacting polarizable particles made of charged point-mass subparticles (partons). He associated the Newtonian gravitational field with the Van Der Waals force present in neutral matter, where the long-range radiation fields are generated by the parton 'Zitterbewegung'. Sakharov went on to develop what he called the 'metric elasticity' concept, where space-time is somehow identified with the 'hydrodynamic elasticity' of the vacuum. However, he did not understand the important clues offered by the equivalence principle, nor the role that the quantum vacuum played in inertia and Mach's principle.

In 1974, Hawkings (ref. 17) announced that black holes are not completely black. Black holes emit an outgoing thermal flux of radiation due to gravitational interactions of the black hole with the virtual particle pairs created in the quantum vacuum near the event horizon. At first sight, the emission of thermal radiation from a black hole seems paradoxical (since nothing can escape from the event horizon). However, the spontaneous creation of virtual particle and anti-particle pairs in the quantum vacuum near the event horizon can be used to explain this effect (ref. 18). Heuristically, one can imagine that the virtual particle pairs (created with wavelength λ, approximately equal to the size of the black hole) 'tunnel' out of the event horizon. For virtual particle with wavelength comparable to the size of the hole, strong tidal forces operate to prevent re-annihilation. One virtual particle escapes to infinity with positive energy to contribute to the Hawking radiation, while the corresponding antiparticle enters the black hole to be trapped forever by the deep gravitational potential. Thus, the quantum vacuum is important in order to understand Hawking Radiation. In EMQG, the quantum vacuum plays an *extremely* important role in both inertia and gravitation. Anybody who believes in the existence of the virtual particles of the quantum vacuum, and accepts the fact that many of them carry mass (virtual fermions in particular) will have no trouble in believing that they are falling in the presence of a large gravitational mass. The existence of the downward accelerating virtual particles (during their brief lifetime) under gravitational fields turns out to be a central theme of EMQG. This idea turns out to be the missing link between inertia and gravity, and leads us directly to a full understanding of the principle of equivalence.

**EMQG and the Quantum Theory of Inertia**

EMQG theory presents a unified approach to Inertia, Gravity, the Principle of Equivalence, Space-Time Curvature, Gravitational Waves, and Mach's Principle. These apparently different phenomena are the common results of the quantum interactions of the real (charged) matter particles (of a mass) with the surrounding virtual particles of the quantum vacuum through the exchange of two force particles: the photon and the graviton. Furthermore, the problem of the cosmological constant is solved automatically in the framework of EMQG. This new approach to quantum gravity is definitely *non-geometric* on the tiniest of distance scales (Plank Scales of distance and time). This is because the large scale relativistic 4D space-time curvature is caused purely by the accelerated state of virtual particles of the quantum vacuum with respect to a mass, and their discrete interactions with real matter particles of a mass through the particle force exchange process. Because of this departure from a universe with fundamentally curved space-time, EMQG is a complete change in paradigm over conventional gravitational physics. This paper should be considered as a framework, or outline of a new approach to gravitational physics that will hopefully lead to a full theory of quantum gravity.

We modified the HRP theory of Inertia (ref. 5) based on our detailed examination of the principle of equivalence. In EMQG, the modified version of inertia is known as "Quantum Inertia", or QI. In EMQG, a new elementary particle is required to fully understand inertia, gravitation, and the principle of equivalence. All matter, including electrons and quarks, must be made of nature's most fundamental mass unit or particle which we call the 'masseon' particle. These particles contain one fixed, fundamental 'quanta' of both inertial and gravitational mass. The masseons also carry one basic, smallest unit or quanta of electrical charge as well, of which they can be either positive or negative. Masseons exist in the particle or in the anti-particle form (called anti-masseon), that can appear at random in the vacuum as virtual masseon/anti-masseon particle pairs of opposite electric charge and opposite 'mass charge'. The earth consists of ordinary masseons (with no anti-masseons), of which there are equal numbers of positive and negative electric charge varieties. In HRP Inertia theory, the electrically charged 'parton' particles (that make up an inertial mass in an accelerated reference frame) interact with the background vacuum electromagnetic zero-point-field (or virtual photons) creating a resistance to acceleration called inertia. We have modified this slightly by postulating that the real masseons (that make up an accelerating mass) interacts with the surrounding, virtual masseons of the quantum vacuum, electromagnetically (although the details of this process are still not fully understood). The properties of the masseon particle and gravitons are developed later.

## EMQG and the Quantum Origin of Newton's Laws of Motion

We are now in a position to understand the quantum nature of Newton's classical laws of motion. According to the standard textbooks of physics (ref. 19) Newton's three laws of laws of motion are:

(1) An object at rest will remain at rest and an object in motion will continue in motion with a constant velocity unless it experiences a net external force.
(2) The acceleration of an object is directly proportional to the resultant force acting on it and inversely proportional to its mass. Mathematically: $\Sigma F = ma$, where F and a are vectors.
(3) If two bodies interact, the force exerted on body 1 by body 2 is equal to and opposite the force exerted on body 2 by body 1. Mathematically: $F_{12} = -F_{21}$.

Newton's first law explains what happens to a mass when the resultant of all external forces on it is zero. Newton's second law explains what happens to a mass when there is a nonzero resultant force acting on it. Newton's third law tells us that forces always come in pairs. In other words, a single isolated force cannot exist. The force that body 1 exerts on body 2 is called the action force, and the force of body 2 on body 1 is called the reaction force.

In the framework of EMQG theory, Newton's first two laws are the direct consequence of the (electromagnetic) force interaction of the (charged) elementary particles of the mass interacting with the (charged) virtual particles of the quantum vacuum. Newton's third law of motion is the direct consequence of the fact that all forces are the end result of a boson particle exchange process.

**NEWTON'S FIRST LAW OF MOTION:**

In EMQG, the first law is a trivial result, which follows directly from the quantum principle of inertia (postulate #3). First a mass is at relative rest with respect to an observer in deep space. If no external forces act on the mass, the (charged) elementary particles that make up the mass maintain a *net acceleration* of zero with respect to the (charged) virtual particles of the quantum vacuum through the electromagnetic force exchange process. This means that no change in velocity is possible (zero acceleration) and the mass remains at rest. Secondly, a mass has some given constant velocity with respect to an observer in deep space. If no external forces act on the mass, the (charged) elementary particles that make up the mass also maintain a *net acceleration* of zero with respect to the (charged) virtual particles of the quantum vacuum through the electromagnetic force exchange process. Again, no change in velocity is possible (zero acceleration) and the mass remains at the same constant velocity.

**NEWTON'S SECOND LAW OF MOTION:**

In EMQG, the second law is the quantum theory of inertia discussed above. Basically the state of *relative* acceleration of the charged virtual particles of the quantum vacuum with respect to the charged particles of the mass is what is responsible for the inertial force. By this we mean that it is the tiny (electromagnetic) force contributed by each mass particle undergoing an acceleration 'A', with respect to the net statistical average of the virtual particles of the quantum vacuum, that results in the property of inertia possessed by all masses. The sum of all these tiny (electromagnetic) forces contributed from each charged particle of the mass (from the vacuum) is the source of the total inertial resistance force opposing accelerated motion in Newton's F=MA. Therefore, inertial mass 'M' of a mass simply represents the total resistance to acceleration of all the mass particles.

**NEWTON'S THIRD LAW OF MOTION:**

According to the boson force particle exchange paradigm (originated from QED) all forces (including gravity, as we shall see) result from particle exchanges. Therefore, the force that body 1 exerts on body 2 (called the action force), is the result of the emission of force exchange particles from (the charged particles that make up) body 1, which are readily absorbed by (the charged particles that make up) body 2, resulting in a force acting on body 2. Similarly, the force of body 2 on body 1 (called the reaction force), is the result of the absorption of force exchange particles that are originating from (the charged particles that make up) body 2, and received by (the charged particles that make up) body 1, resulting in a force acting on body 1. An important property of charge is the ability to readily emit and absorb boson force exchange particles. Therefore, body 1 is both an emitter and also an absorber of the force exchange particles. Similarly, body 2 is also both an emitter and an absorber of the force exchange particles. This is the reason that there is both an action and reaction force. For example, the contact forces (the mechanical forces that Newton was thinking of when he formulated this law) that results from a person pushing on a mass (and the reaction force from the mass pushing on the person) is really the exchange of photon particles from the charged electrons bound to the atoms of the person's hand and the charged electrons bound to the atoms of the mass on the quantum level. Therefore, on the quantum level there is really is no contact here. The hand gets very close to the mass, but does not actually touch. The electrons exchange photons among each other. The force exchange process works both directions in equal numbers, because all the electrons in the hand and in the mass are electrically charged and therefore the exchange process gives forces that are equal and opposite in both directions.

**Introduction to the Principle of Equivalence and EMQG**

Are virtual particle force exchange processes originating from the quantum vacuum also present for gravitational mass? The answer turns out to be a resounding yes! As we suggested, there is some evidence of the interplay between the virtual particles of the quantum vacuum and gravitational phenomena. In order to see how this impacts our understanding of the nature of gravitational mass, we found it necessary to perform a thorough investigation of Einstein's Principle of Equivalence of inertial and gravitational mass in general relativity under the guidance of the new theory of quantum inertia.

We have uncovered some theoretical evidence that the SEP may not hold for certain experiments. There are two basic theoretical problems with the SEP in regards to quantum gravity. First, if gravitons (the proposed force exchange particle) can be detected with some new form of a sensitive graviton detector, we would be able to distinguish between an accelerated reference frame and a gravitational frame with this detector. This is because accelerated frames would have virtually no graviton particles present, whereas gravitational fields like the earth have enormous numbers of graviton particles. Secondly, theoretical considerations from several authors (ref. 23) regarding the emission of electromagnetic waves from a uniformly accelerated charge, and the lack of radiation from the same charge subjected to a static gravitational field leads us to question the validity of the SEP for charged particles radiating electromagnetically.

How does the WEP hold out in EMQG? The WEP has been tested to a phenomenal accuracy (ref 24.) in recent times. Yet, in our current understanding of the WEP, we can only specify the accuracy as to which the two different mass values (or types) have been shown experimentally to be equal inside an inertial or gravitational field. There exists no physical or mathematical proof that the WEP is precisely true. It is still only a postulate of general relativity. We have applied the recent work on quantum inertia (ref. 5) to the investigation of the weak principle of equivalence, and have found theoretical reasons to believe that the WEP is not precisely correct when measured in extremely accurate experiments. Imagine an experiment with two masses; one mass $M_1$ being very large in value, and the other mass $M_2$ is very small ($M_1 >> M_2$). These two masses are dropped simultaneously in a uniform gravitational field of 1g from a height 'h', and the same pair of masses are also dropped inside a rocket accelerating at 1g, at the same height 'h'. We predict that there should be a minute deviation in arrival times on the surface of the earth (only) for the two masses, known as the 'Ostoma-Trushyk effect', with the heavier mass arriving just slightly ahead of the smaller mass. This is due to a small deviation in the magnitude of the force of gravity on the mass pair (in favor of $M_1$) on the order of $(N_1-N_2)i * \delta$, where $(N_1-N_2)$ is the difference in the low level mass specified in terms of the difference in the number of masseon particles in the two masses (defined latter) times the single masseon mass 'i', and $\delta$ is the ratio of the gravitational to electromagnetic forces for a single (charged) masseon. This experiment is very difficult to perform on the earth, because $\delta$ is extremely small ($\approx 10^{-40}$), and $\Delta N = (N_1-N_2)$ cannot in practice be made sufficiently large to produce a measurable effect. However, inside the accelerated rocket, the arrival times are <u>exactly</u> identical for the same pair of masses. This, of course, violates the principle of equivalence, since the motion of the masses in the inertial frame is slightly different then in the gravitational frame. This imbalance is minute because of the dominance of the strong electromagnetic force which is also acting on the masseons of the two masses from the virtual particles of the quantum vacuum. This acts to stabilize the fall rate, giving us nearly perfect equivalence.

This conclusion is based on the discovery that the weak principle of equivalence results from lower level physical processes. Mass equivalence arises from the equivalence of the force generated between the net statistical average acceleration vectors of the charged matter particles inside a mass with respect to the immediate surrounding quantum vacuum virtual particles inside an accelerating rocket. This is almost exactly the same physical force occurring between the stationary (charged) matter particles and the immediate surrounding accelerating virtual particles of the same mass near the earth. It turns out that equivalence is not perfect in the presence of a large gravitational field like the earth. Equivalence breaks down due to an extremely minute force imbalance in favor of a larger mass dropped simultaneously with respect to a smaller mass. This force imbalance can be traced to the pure graviton exchange force component occurring in the gravitational field that is not present in the case of the identically dropped masses in an accelerated rocket. This imbalance contributes a minute amount of extra force for the larger mass compared to the smaller mass (due to many more gravitons exchanged between the larger mass as compared to the smaller mass), which might be detected in highly accurate measurements. In the case of the rocket, the equivalence of two different falling masses is perfect, since it is the floor of the rocket that accelerates up to meet the two masses simultaneously. Of course, the breakdown of the WEP also means the downfall of the SEP.

In EMQG, the gravitational interactions involve the same electromagnetic force interaction as found in inertia based on our QI theory. We also found that the weak principle of equivalence itself is a physical phenomenon originating from the hidden lower level quantum processes involving the quantum vacuum particles, graviton exchange particles, and photon exchange particles. In other words, gravitation is purely a quantum force particle exchange process, and is not based on low level fundamental 4D curved space-time geometry of the universe as believed in general relativity. The perceived 4D curvature is a manifestation of the dynamic state of the falling virtual particles of the quantum vacuum in accelerated frames, and gravitational frames. The only difference between the inertial and gravitation force is that gravity also involves graviton exchanges (between the earth and the quantum vacuum virtual particles, which become accelerated downwards), whereas inertia does not. Gravitons have been proposed in the past as the exchange particle for gravitational interactions in a quantum field theory of gravity without much success. The reason for the lack of success is that graviton exchange is not the only exchange process occurring in large-scale gravitational interactions; photon exchanges are also involved! It turns out that not only are there both graviton and photon exchange processes occurring simultaneously in large scale gravitational interactions such as on the earth, but that both exchange particles are almost identical in their fundamental nature (Of course, the strength of the two forces differs greatly).

The equivalence of inertial and gravitational mass is ultimately traced down to the reversal of all the relative acceleration vectors of the charged particles of the accelerated mass <u>with respect</u> to the (net statistical) average acceleration of the quantum vacuum particles, that occurs when changing from inertial to gravitational frames. The inertial mass 'M' of an object with acceleration 'a' (in a rocket traveling in deep space, away from gravitational fields) results from the sum of all the tiny forces of the charged elementary particles that make up that mass with respect to the immediate quantum vacuum particles. This inertial force is in the opposite direction to the motion of the rocket. The (charged masseon) particles building up the mass in the rocket will have a net statistical average acceleration 'a' with respect to the local (charged masseon) virtual particles of the immediate quantum vacuum. A stationary gravitational mass resting on the earth's surface has this same quantum process occurring as for the accelerated mass, but with the acceleration vectors reversed. What we mean by this is that under gravity, it is now the virtual particles of the quantum vacuum that carries the net statistical average acceleration 'A' downward. This downward virtual particle acceleration is caused by the graviton exchanges between the earth and the mass, where the mass is not accelerated with respect to the center of mass of the earth. (Note: On an individual basis, the velocity vectors of these quantum vacuum particles actually point in all directions, and also have random amplitudes. Furthermore, random accelerations occur due to force interactions between the virtual particles themselves. This is why we refer to the statistical nature of the acceleration.) We now see that the gravitational force of a stationary mass is also the <u>same</u> sum of the tiny forces that originate for a mass undergoing accelerated motion in a gravitational field from the virtual particles of the quantum vacuum according to Newton's law 'F = MA'. In other words, the same inertial force F=MA is also found hidden inside gravitational interactions of masses! Mathematically, this fact can be seen in Newton's laws of inertia and in Newton's gravitational force law by slightly rearranging the formulas as follows:

$F_i = M_i (A_i)$ ... the inertial force $F_i$ opposes the acceleration $A_i$ of mass $M_i$ in the rocket, caused by the sum of the tiny forces from the virtual particles of the quantum vacuum.

$F_g = M_g (A_g) = M_g (GM_e/r^2)$ ... the gravitational force $F_g$ is the result of a kind of an inertial force given by '$M_g A_g$' where $A_g = GM_e/r^2$ is now due to the sum of the tiny forces from the virtual particles of the quantum vacuum (now accelerating downwards).

Since $F_i = F_g$, and since the acceleration of gravity is chosen to be the same as the inertial acceleration, where the virtual particles now have: $A_g = A_i = GM_e/r^2$, therefore $M_i = M_g$, or the inertia mass is equal to the gravitational mass ($M_e$ is the mass of the earth). Here, Newton's law of gravity is rearranged slightly to emphasis it's form as a kind of an 'inertial force' of the form F=MA, where the acceleration ($GM_e/r^2$) is now the net statistical average downward acceleration of the quantum vacuum virtual particles near the vicinity of the earth.

This derivation is not complete, unless we can provide a clear explanation as to why $F_i = F_g$, which we know to be true from experimental observation. In EMQG, both of these forces are understood to arise from an almost identical quantum vacuum process! For accelerated masses, inertia is the force $F_i$ caused by the sum of all the tiny electromagnetic forces from each of the accelerated charged particles inside the mass; with respect to the non-accelerating surrounding virtual particles of the quantum vacuum. Under the influence of a gravitational field, the <u>same force</u> $F_g$ exists as it does in inertia, but now the quantum vacuum particles are the ones undergoing the same acceleration $A_i$ (through graviton exchanges with the earth); the charged particles of the mass are stationary with respect to earth's center. The same force arises, but the arrows of the acceleration vectors are reversed. To elaborate on this, imagine that you are in the reference frame of a stationary mass resting on the surface with respect to earth's center. An average charged particle of this mass 'sees' the virtual particles of the quantum vacuum in the same state of acceleration, as does an average charged particle of an identical mass sitting on the floor of an accelerated rocket (1 g). In other words, the background quantum vacuum 'looks' exactly the same from both points of view (neglecting the very small imbalance caused by a very large number of gravitons interacting with the mass directly under gravity, this imbalance is swamped by the strength of the electromagnetic forces existing).

These equations and methodology illustrates equivalence in a special case: i.e., between an accelerated mass $M_i$ and the same stationary gravitational mass $M_g$. In EMQG, the weak equivalence principle of gravitational and inertial frames holds for many other scenarios such as for free falling masses, for masses that have considerable self gravity and energy (like the earth), for elementary particles, and for the propagation of light. However, equivalence is *not* perfect, and in some situations (for example, antimatter discussed in section 7.1) it simply does not hold at all!

An astute observer may question why all the virtual particles (electrons, quarks, etc, all having different masses) are accelerating downwards on the earth with the same acceleration. This definitely would be the case from the perspective of a mass being accelerated by a rocket (where the observer is accelerating). Since the masses of the different types of virtual particles are all different according to the standard model of particle physics, why are they all falling at the same rate? Since we are trying to derive the equivalence principle, we cannot invoke this principle to state that all virtual particles must be accelerating downward at the same rate. It turns out that the all quantum vacuum virtual particles are accelerating at the same rate because all particles with mass (virtual or not) are composed of combinations of a new fundamental "masseon" particle which carries just one fixed quanta of mass. Therefore, all the elementary virtual masseon particles of the quantum vacuum are accelerated by the same amount. These masseons can bind together to form the familiar particles of the standard model, like virtual electrons, virtual positrons, virtual quarks, etc. Recalling that the masseon also carries electrical charge, we see that all the constituent masseons of the quantum vacuum particles fall to earth at same rate through the electromagnetic interaction (or photon exchange) process, no matter how the virtual masseons combine to give the familiar virtual particles. This process works like a microscopic principle of equivalence for falling virtual particles, with the same action occurring for virtual particles as for large falling masses.

The properties of the masseon particle is elaborated in section 7 (the masseon may be the unification particle sought out by physicist, in which case it will have other properties to do with the other forces of nature). For now, note that the masseon also carries the fundamental unit of electric charge as well. This fundamental unit of electric charge turns out to be the source of inertia for all matter according to Quantum Inertia. By postulating the existence of the masseon particle (which is the fundamental unit of 'mass charge' as well as 'electrical charge') all the quantum vacuum virtual particles accelerate at the same rate with respect to an observer on the surface of the earth. We have postulated the existence of a fundamental "low level gravitational mass charge" of a particle, which results from the graviton particle exchange process similar to the process found for electrical charges. This 'mass charge' is not affected when a particle achieves relativistic velocities, so we can state that 'low level mass charge' is an absolute constant. For particles accelerated to relativistic speeds, a high relative velocity between the source of the force and the receiving mass affects the ordinary measurable inertial mass, as we have seen (in accordance to Einstein's mass-velocity formula).

**Summary of the Basic Mass Definitions in EMQG**

EMQG proposes three different mass definitions for an object:

(1) INERTIAL MASS is the measurable mass defined in Newton's force law F=MA. This is considered as the absolute mass in EMQG, because it results from force produced by the relative (statistical average) acceleration of the charged virtual particles of the quantum vacuum with respect to the charged particles that make up the inertial mass. The virtual particles of the quantum vacuum form Newton's absolute reference frame. In special relativity this mass is equivalent to the rest mass.

(2) GRAVITATIONAL MASS is the measurable mass involved in the gravitational force as defined in Newton's law $F=GM_1M_2/R^2$. This is what is measured on a weighing scale. This is also considered as absolute mass, and is almost exactly the same as inertial mass.

(3) LOW LEVEL GRAVITATIONAL 'MASS CHARGE' which is the origin of the pure gravitational force, is defined as the force that results through the exchange of graviton particles between two (or more) quantum particles. This type of mass analogous to 'electrical charge', where photon particles are exchanged between electrically charged particles. Note: this force is very hard to measure because it is masked by the background quantum vacuum electromagnetic force interactions, which dominates over the graviton force processes.

These three forms of mass are <u>not</u> necessarily equal! We have seen that the inertial mass is almost exactly the same as gravitational mass, but not perfectly equal. All quantum mass particles (fermions) have all three mass types defined above. But bosons (only photons and gravitons are considered here) have only the first two mass types. This means that photons and gravitons transfer momentum, and <u>do</u> react to the presence of inertial frames and to gravitational fields, but they do not emit or absorb gravitons. Gravitational fields affect photons, and this is linked to the concept of space-time curvature, described in detail later (section 9). It is important to realize that gravitational fields deflect photons (and gravitons), but not by force particle exchanges directly. Instead, it is due to a scattering process (described later).

To summarize, both the photon and the graviton do not carry low level 'mass charge', even though they both carry inertial and gravitational mass. The graviton exchange particle, although responsible for a major part of the gravitational mass process, does not itself carry the property of 'mass charge'. Contrast this to conventional physics, where the photon and the graviton both carry a non-zero mass given by $M=E/C^2$. According to this reasoning, the photon and the graviton both carry mass (since they carry energy), and therefore both must have 'mass charge' and exchange gravitons. In other words, the graviton particle not only participates in the exchange process, it also undergoes further exchanges while it is being exchanged! This is the source of great difficulty for canonical quantum gravity theories, and causes all sorts of mathematical renormalization problems in the corresponding quantum field theory. Furthermore, in gravitational force interactions with photons, the strength of the force (which depends on the number of gravitons exchanged with photon) varies with the energy that the photon carries! In modern physics, we do not distinguish between inertial, gravitational, or low level 'mass charge'. They are assumed to be equivalent, and given a generic name 'mass'. In EMQG, the photon and graviton carry measurable inertial and gravitational mass, but neither particle carries the 'low level mass charge', and therefore do not participate in graviton exchanges.

We must emphasize that gravitons do not interact with each other through force exchanges in EMQG, just as photons do not interact with each other with force exchanges in QED. Imagine if gravitons did interact with other gravitons. One might ask how it is possible for a graviton particle (that always moves at the speed of light) to emit graviton particles that are also moving at the speed of light. For one thing, this violates the principles of special relativity theory. Imagine two gravitons moving in the same direction at the speed of light that are separated by a distance d, with the leading graviton called 'A' and the lagging graviton called 'B'. How can graviton 'B' emit another graviton (also moving at the speed of light) that gets absorbed by graviton 'A' moving at the speed of light? As we have seen, these difficulties are resolved by realizing that there are actually three different types of mass. There is measurable inertial mass and measurable gravitational mass, and low level 'mass charge' that cannot be directly measured. Inertial and gravitational mass have already been discussed and arise from different physical circumstances, but in most cases give identical results. However, the 'low level mass charge' of a particle is defined simply as the force existing between two identical particles due to the exchange of graviton particles only, which are the vector bosons of the gravitational force. Low level mass charge is not directly measurable, because of the complications due to the electromagnetic forces simultaneously present from the quantum vacuum virtual particles.

It would be interesting to speculate what the universe might be like if there were no quantum vacuum virtual particles present. Bearing in mind that the graviton exchange process is almost identical to the photon exchange process, and bearing in mind the complete absence of the electromagnetic component in gravitational interactions, the universe would be a very strange place. We would find that large masses would fall faster than smaller masses, just as a large positive electric charge would 'fall' faster then a small positive charge towards a very large negative charge. There would be no inertia as we know it, and basically no force would be required to accelerate or stop a large mass.

## The Quantum Field Theory of the Masseon and Graviton Particles

EMQG addresses gravitational force, inertia, and electromagnetic forces only, and the weak and strong nuclear forces are excluded from consideration. EMQG is based on the idea that all elementary matter particles must get their quantum mass numbers from combinations of just one fundamental matter (and corresponding anti-matter particle), which has just one fixed unit or quanta of mass which we call the 'masseon' particle. This fundamental particle generates a fixed flux of gravitons that are exchanged during gravitational interactions. The exchange process is not affected by the state of motion of the masseon (as you might expect from the special relativistic variation of mass with velocity). We also purpose that nature does <u>not</u> have two completely different long-range forces, e.g. gravity and electromagnetism. Instead, we believe that there exists an almost perfect symmetry between the two forces, which is hidden from view because of the mixing of these two forces in all measurable gravitational interactions. In EMQG, the graviton and photon exchange process is found to be essentially the same, except for the strength of the force coupling (and a minor difference in the treatment of positive and negative masses discussed later). EMQG treats graviton exchanges by the same successful methods developed for the behavior of photons in QED. The dimensionless coupling constant that governs the graviton exchange process is what we call '$\beta$' in close analogy with the dimensionless coupling constant '$\alpha$' in QED, where $\beta \approx 10^{-40} \alpha$.

As we stated, EMQG requires the existence of a new fundamental matter particle called the 'masseon' (and a corresponding 'anti-masseon' particle), which are held together by a new unidentified strong force. Furthermore, EMQG requires that masseons and anti-masseons emit gravitons analogous with the electrons and anti-electrons (positrons) which emit photons in QED. Virtual masseons and anti-masseons are created in equal amounts in the quantum vacuum as virtual particle pairs. A masseon generates a fixed flux of graviton particles with wave functions that induce attraction when absorbed by another masseon or anti-masseon; and an anti-masseon generates a fixed flux of graviton particles with an opposite wave function (anti-gravitons) that induces repulsion when absorbed by another masseon or anti-masseon. A graviton is its own anti-particle, just as a photon is its own antiparticle. This process is similar to, but not identical to the photon exchange processes in QED for electrons of opposite charge. In QED, an electron produces a fixed flux of photon particles with wave functions that induce repulsion when absorbed by another electron, and induces attraction when absorbed by a positron. A positron produces a fixed flux of photon particles with wave functions that induce attraction when absorbed by another electron, and induces repulsion when absorbed by a positron. From this it can be seen that if two sufficiently large pieces of anti-matter can be fabricated which are both electrically neutral, they will be found to repel each other gravitationally! Thus, anti-matter can be thought of as literally 'negative' mass (-M), and therefore negative energy. This grossly violates the equivalence principle.

These subtle differences in the exchange process in QED and EMQG produce some interesting effects for gravitation that are not found in electromagnetism. For example, a large gravitational mass like the earth does not produce vacuum polarization of virtual particles from the point of view of 'mass-charge' (unlike electromagnetism). In gravitational fields, all the virtual masseon and anti-masseon particles of the vacuum have a net average statistical acceleration directed downwards towards a large mass. This produces a net downward accelerated flux of vacuum particles (acceleration vectors only) that affects other masses immersed in this flux.

In contrast to this, an electrically charged object <u>does</u> produce vacuum polarization. For example, a negatively charged object will cause the positive and negative (electrically charged) virtual particles to accelerate towards and away, respectively from the negatively charged object. Therefore, there is no energy contribution to other real electrically charged test particles placed near the charged object from the vacuum particles, because the electrically charged vacuum particles contributes equal amounts of force contributions from both the upward and downward directions. The electrical forces from the vacuum cancels out to zero.

In gravitational fields, the vacuum particles are responsible for the principle of equivalence, precisely because of the lack of vacuum polarization due to gravitational fields. Recall that 'masseon' particles of EMQG are equivalent to the 'parton' particle concept that was introduced by the authors of reference 5 in regards to HRP Quantum Inertia. Recall that the masseons and anti-masseons also carry one quanta of electric charge of which there are two types; positive and negative charges. For example, masseons come in positive and negative electric charge, and anti-masseons also come in positive and negative charge. A single charged masseon particle accelerating at 1g sees a certain fixed amount of inertial force generated by the virtual particles of the quantum vacuum. In a gravitational field of 1g, a single charged masseon particle on the surface of the earth sees the same quantum vacuum electromagnetic force. In other words, from the vantage point of a masseon particle that makes up the total mass, the virtual particles of the quantum vacuum looks exactly the same from the point of view of motion and forces whether it is in an inertial reference frame or in a gravitational field. We propose a new universal constant "i" called the 'inertion', which is defined as the inertial force produced by the action of virtual particles on a single (real charged) masseon particle undergoing a relative acceleration of 1 g. This force is the lowest possible quanta of inertial mass. All other masses are fixed integer combinations of this number. This same constant 'i' is also the lowest possible quanta of gravitational force.

The electric charge that is carried by the electron, positron, quark and anti-quark originates from combinations of masseons, which is the fundamental source of the electrical charge. This explains why a fixed charge relationship exists between the quarks and the leptons, which belong to different families in the standard model. For example, according to the standard model, 1 proton charge precisely equals 1 electron charge (opposite polarity), where the proton is made of 3 quarks. This precise equality arises from the fact that charged masseon particles are present in the internal structure of both the quarks and the electrons (and every other mass particle).

The mathematical renormalization process is applied to particles to avoid infinities encountered in Quantum Field Theory (QFT) calculations. This is justified by postulating a high frequency cutoff of the vacuum processes in the summation of the Feynman diagrams. Recall that QED is formulated on the assumption that a perfect space-time continuum exists. In EMQG, a high frequency cutoff is essential because space is quantized as 'cells', specified by Cellular Automata (CA) theory. In CA theory there is quantization of space in the form of cells. If particles are sufficiently close enough, they completely lose their identity as particles in CA theory, and QFT does not apply at this scale. Since graviton exchanges are almost identical to photon exchanges, we suspect that EMQG is also renormalizable as is QED, with a high frequency cutoff as well. This has not been proven yet. The reason that some current quantum gravity theories are not renormalizable boils down to the fact that the graviton is assumed to be the only boson involved in gravitational interactions. The graviton must therefore exhibit all the characteristics of the gravitational field, including space-time curvature.

In EMQG, the photon exchange and graviton exchange process is virtually identical in its basic nature, which shows the great symmetry between these two forces. As a byproduct of this, the quantum vacuum becomes 'neutral' in terms of gravitational 'mass charge', as the quantum vacuum is known to be neutral with respect to electric charge. This is due to an equal number of positive and negative electrical charged virtual particles and 'gravitational charged' virtual particles created in the quantum vacuum at any given time. This in turn is due to the symmetrical masseon and anti-masseon pair creation process. (EMQG does not resolve the problem of why the universe was created with an apparent imbalance of real ordinary matter and anti-matter mass particles.)

This distortion of the acceleration vectors of the quantum vacuum 'stream' serves as an effective 'electromagnetic guide' for the motion of nearby test masses (themselves consisting of masseons) through space and time. This 'electromagnetic guide' concept replaces the 4D space-time geodesics (which is the path that light takes through curved 4D space-time) that guide light and matter in motion. Because the quantum vacuum virtual particle density is quite high, but not infinite (at least about $10^{90}$ particles/m$^3$), the quantum vacuum acts as a very effective and energetic guide for the motion of light and matter.

### Introduction to 4D Space-Time Curvature and EMQG

The physicist A. Wheeler once said that: "space-time geometry 'tells' mass-energy how to move, and mass-energy 'tells' space-time geometry how to curve". In EMQG, this statement must be somewhat revised on the quantum particle level to read: large mass-energy concentrations (consisting of quantum particles) exchanges gravitons with the immediate surrounding virtual particles of the quantum vacuum, causing a downward acceleration (of the net statistical average acceleration vectors) of the quantum vacuum particles. This downward acceleration of the virtual particles of the quantum vacuum 'tells' a nearby test mass (also consisting of real quantum particles) how to move electromagnetically, by the exchange of photons between the electrically charged, and falling virtual particles of the quantum vacuum and the electrical charged, real particles inside the test mass. This new view of gravity is totally based on the ideas of quantum field theory, and thus acknowledging the true particle nature of both matter and forces. It is also shows how nature is non-geometric when examined on the smallest of distance scales, where Riemann geometry is now replaced solely by the interactions of quantum particles existing on a kind of quantized 3D space and separate time on the CA.

Since this downward accelerated stream of charged virtual particles also affects light or real photons and the motion of real matter (for example, matter making up a clock), the concept of space-time must be revised. For example, a light beam moving parallel to the surface of the earth is affected by the downward acceleration of charged virtual particles (electromagnetically), and moves in a curved path. Since light is at the foundation of the measurement process as Einstein showed in special relativity, the concept of space-time must also be affected near the earth by this accelerated 'stream' of virtual particles. Nothing escapes this 'flow', and one can imagine that not even a clock is expected to keep the same time as it would in far space. As a result, a radically new picture of Einstein's curved space-time concept arises from these considerations in EMQG.

The variation of the value of the net statistical average (directional) acceleration vector of the quantum vacuum particles from point to point in space (with respect to the center of a massive object) guides the motion of nearby test masses and the motion of light through electromagnetic means. This process leads to the 4D space-time metric curvature concept of general relativity. With this new viewpoint, it is now easy to understand how one can switch between accelerated and gravitational reference frames. Gravity can be made to cancel out inside a free falling frame (technically at a point) above the earth because we are simply taking on the same net acceleration as the virtual particles at that point. In this scenario, the falling reference frame creates the same quantum vacuum particle background environment as found in an non-accelerated frame, far from all gravitational fields. As a result, light travels in perfectly straight lines when viewed by a falling observer, as specified by special relativity. Thus, in the falling reference frame, a mass 'feels' no force or curvature as it would in empty space, and light travels in straight lines (defined as 'flat' space-time). Thus, the mystery as to why different reference frames produce different space-time curvature is solved in EMQG. It is interesting that in an accelerated rocket, space-time curvature is also present, but is now caused by another mechanism; the accelerated motion of the floor of the rocket itself. In other words, the space-time curvature, manifesting itself as the path of curved light, is really caused by the accelerated motion of the observer! The observer (now in a state of acceleration with respect to the vacuum), 'sees' the accelerated virtual particle motion in his frame. Furthermore, the motion appears to him to be almost exactly the same as if he were in an equivalent gravitational field. This is why the space-time curvature appears the same in both a gravitational field and an equivalent accelerated frame. These differences between accelerated and gravitational frames imply that equivalence is not a basic element of reality, but merely a result of different physical processes, which happen to give the same results. In fact, equivalence is <u>not</u> perfect!

According to EMQG, all metric theories of gravity, including general relativity, have a limited range of application. These theories are useful only when a sufficient mass is available to significantly distort the virtual particle motion surrounding the mass; and only where the electromagnetic interactions dominates over the graviton processes (or where the graviton flux is not too large). For precise calculation of gravitational force interactions of small masses, EMQG requires that the gravitational interaction be calculated by adding the specific Feynman diagrams for both photon and graviton exchanges. Thus, the use of the general relativistic Schwarzchild Metric for spherical bodies (even if modified by including the uncertainty principle) is totally useless for understanding the gravitational interactions of elementary particles. The whole concept of space-time 'foam' is incorrect according to EMQG, along with all the causality problems associated with this complex mathematical concept.

**Space-Time Curvature is a Pure Virtual Particle Quantum Vacuum Process**

4D Minkowski curved space-time takes on a radically new meaning in EMQG, and is no longer a basic physical element of our reality. Instead, it is merely the result of quantum particle interactions alone. The curved space-time of general relativity arises strictly out of the interactions between the falling virtual particles of the quantum vacuum near a massive object and a nearby test mass. The effect of the falling quantum vacuum acts somewhat like a special kind of "Fizeau-Fluid" or media, that affects the propagation of light; and also effects clocks, rulers, and measuring instruments. Fizeau demonstrated in the middle 1850's that moving water varies the velocity of light propagating through it. This effect was analyzed mathematically by Lorentz. He used his newly developed microscopic theory for the propagation of light in matter to study how photons move in a flowing stream of transparent fluid. He reasoned that photons would change velocity by frequent scattering with the molecules of the water, where the photons are absorbed and later remitted after a small time delay. This concept is discussed in detail in section 9.3. If Einstein himself had known about the existence of the quantum vacuum when he was developing general relativity theory, he may have deduced that space-time curvature was caused by the "accelerated quantum vacuum fluid". He was aware of the work by Fizeau, but was unaware of the existence of the quantum vacuum. After all, Einstein certainly realized that clocks were not expected to keep time correctly when immersed in an accelerated stream of water! We show mathematically in this paper that the quantity of space-time curvature near a spherical object predicted by the Schwarzchild metric is identical to the value given by the 'Fizeau-like' scattering process in EMQG.

In EMQG, anywhere we find accelerated vacuum disturbance, there follows a corresponding space-time distortion (including gravitational waves). We have seen that both accelerated and gravitational frames qualify for the status of curved 4D space-time (although caused by <u>different</u> physical circumstances). We have found that in EMQG there exists two, separate but related space-time coordinate systems. First, there is the familiar global four dimensional relativistic space-time of Minkowski, as defined by our measuring instruments, and is designated by the x,y,z,t in Cartesian coordinates. The amount of 4D space-time curvature is influenced by accelerated frames and by gravitational frames, which is the cause of the accelerated state of the quantum vacuum.

Secondly there is a kind of a quantized absolute space, and separate time as required by cellular automata theory. Absolute space consists of an array of numbers or cells C(x,y,z) that changes state after every new clock operation $\Delta t$. C(x,y,z) acts like the absolute three dimensional pre-relativistic space, with a separate absolute time that acts to evolve the numerical state of the cellular automata. The CA space (and separate time) is not effected by any physical interactions or directly accessible through any measuring instruments, and currently remains a postulate of EMQG. Note that EMQG absolute space does not correspond to Newton's idea of absolute space. Newton postulated the existence of absolute space in his work on inertia. He realized that absolute space was required in order to resolve the puzzle of what reference frame nature uses to gauge accelerated motion. In EMQG, this reference frame is <u>not</u> the absolute quantized cell space (which is unobservable), but instead consists of the net average state of acceleration of the virtual particles of the quantum vacuum with respect to matter (particles). A very important consequence of the existence of absolute quantized space and quantized time (required by cellular automata theory) is the fact that our universe must have a maximum speed limit!

## 4. THE VIRTUAL PARTICLES OF THE QUANTUM VACUUM

*Philosophers:     "Nature abhors a vacuum."*

Because of the central importance of the virtual particles of the quantum vacuum to understanding the principle of equivalence, we present a brief review of the development of the quantum vacuum concept and the experimental evidence for it's existence. One might think that the vacuum is completely devoid of everything. In fact, the vacuum is far from empty. In order to make a complete vacuum, one must remove all matter from an enclosure. However, this is still not good enough. One must also lower the temperature down to absolute zero in order to remove all thermal electromagnetic radiation. However, Nernst correctly deduced in 1916 (ref. 32) that empty space is still not completely devoid of all radiation after this is done. He predicted that the vacuum is still permanently filled with an electromagnetic field propagating at the speed of light, called the zero-point fluctuations (sometimes called vacuum fluctuations). This was later confirmed by the full quantum field theory developed in the 1920's and 30's. Later, with the development of QED, it was realized that all quantum fields should contribute to the vacuum state, like virtual electrons and positron particles, for example.

According to modern quantum field theory, the perfect vacuum is teeming with activity, as all types of quantum virtual particles (and virtual bosons or force particles) from the various quantum fields, appear and disappear spontaneously. These particles are called 'virtual' particles because they result from quantum processes that have short lifetimes, and are undetectable.

One way to look at the existence of the quantum vacuum is to consider that quantum theory forbids the absence of motion, as well as the absence of propagating fields (exchange particles). In QED, the quantum vacuum consists of the virtual particle pair creation/annihilation processes (for example, electron-positron pairs), and the zero-point-fluctuation (ZPF) of the electromagnetic field (virtual photons) just discussed. The existence of virtual particles of the quantum vacuum is essential to understanding the famous Casimir effect (ref. 11), an effect predicted theoretically by the Dutch scientist Hendrik Casimir in 1948. The Casimir effect refers to the tiny attractive force that occurs between two neutral metal plates suspended in a vacuum. He predicted theoretically that the force 'F' per unit area 'A' for plate separation D is given by:

$$F/A = -\pi^2 h c /(240 D^4) \quad \text{Newton's per square meter} \quad \text{(Casimir Force 'F')} \quad (4.1)$$

The origin of this minute force can be traced to the disruption of the normal quantum vacuum virtual photon distribution between two nearby metallic plates. Certain photon wavelengths (and therefore energies) in the low wavelength range are not allowed between the plates, because these waves do not 'fit'. This creates a negative pressure due to the unequal energy distribution of virtual photons inside the plates as compared to outside the plate region. The pressure imbalance can be visualized as causing the two plates to be drawn together by radiation pressure. Note that even in the vacuum state, virtual photons carry energy and momentum.

Recently, Lamoreaux made (ref. 12) accurate measurements for the first time on the theoretical Casimir force existing between two gold-coated quartz surfaces that were spaced 0.75 micrometers apart. Lamoreaux found a force value of about 1 billionth of a Newton, agreeing with the Casimir theory to within an accuracy of about 5%.

EMQG theory depends heavily on the existence of the virtual particles of the quantum vacuum, and so we present other evidence for the existence of virtual particles (briefly) below:

(1) The extreme precision in the theoretical calculations of the hyper-fine structure of the energy levels of the hydrogen atom, and the anomalous magnetic moment of the electron and muon are both based on the existence of virtual particles. These effects have been calculated in QED to a very high precision (approximately 10 decimal places), and these values have also been verified experimentally. This indeed is a great achievement for QED, which is essentially a perturbation theory of the electromagnetic quantum vacuum.

(2) Recently, vacuum polarization (the polarization of electron-positron pairs near a real electron particle) has been observed experimentally by a team of physicists led by David Koltick (ref. 33). Vacuum polarization causes a cloud of virtual particles to form around the electron in such a way as to produce charge screening. This is because virtual positrons migrate towards the real electron and virtual electrons migrate away. A team of physicists fired high-energy particles at electrons, and found that the effect of this cloud of virtual particles was reduced, the closer a particle penetrated towards the electron. They reported that the effect of the higher charge for the penetration of the electron cloud with energetic 58 giga-electron volt particles was equivalent to a fine structure constant of 1/129.6. This agreed well with their theoretical prediction of 128.5. This can be taken as verification of the vacuum polarization effect predicted by QED.

(3) The quantum vacuum explains why cooling alone will never freeze liquid helium. Unless pressure is applied, vacuum energy fluctuations prevent its atoms from getting close enough to trigger solidification.

(4) For fluorescent strip lamps, the random energy fluctuations of the vacuum cause the atoms of mercury, which are in their exited state, to spontaneously emit photons by eventually knocking them out of their unstable energy orbital. In this way, spontaneous emission in an atom can be viewed as being caused by the surrounding quantum vacuum.

(5) In electronics, there is a limit as to how much a radio signal can be amplified. Random noise signals are somehow added to the original signal. This is due to the presence of the virtual particles of the quantum vacuum as the photons propagate in space, thus adding a random noise pattern to the signal.

(6) Recent theoretical and experimental work in the field of Cavity Quantum Electrodynamics suggests that orbital electron transition time for excited atoms can be affected by the state of the virtual particles of the quantum vacuum surrounding the excited atom in a cavity.

## 5. GENERAL RELATIVITY, ACCELERATION, AND GRAVITY

**"The general laws of physics (and gravitation) are to be expressed by equations which hold good for all systems of coordinates."**

**- Albert Einstein**

Einstein's gravitational field equations are a set of observer dependent equations for observers that are subjected to gravity and/or to acceleration. These equations are based on *measurable* 4D space-time. The core of Einstein's theory is the principle of equivalence and the principle of general covariance, which allow an observer in any state of motion (and coordinate system) to describe gravity and acceleration. However, CA theory places little significance to an observer unless the observer interferes with the interaction being measured. In a CA, physical processes continue without regards to the presence of an observer, where events unfold in absolute space and time.

We now briefly review the general theory of relativity.

**POSTULATES OF GENERAL RELATIVITY**

General relativity is a classical field theory founded on all the postulates and results of special relativity, as well as on the following new postulates:

**(1) PRINCIPLE OF EQUIVALENCE (STRONG) - The results of any given physical experiment will be precisely identical for an accelerated observer in free space as it is for a non-accelerated observer in a perfectly uniform gravitational field. A weaker form of this postulate states that: objects of the different mass fall at the same rate of acceleration in a uniform gravity field.**

**(2) PRINCIPLE OF COVARIANCE - The general laws of physics can be expressed in a form that is independent of the choice of space-time coordinates and the state of motion of an observer.**

As a consequence of postulate 1, the inertial mass of an object is equivalent to it's gravitational mass. Einstein uses this principle to encompass gravity and inertia into his single framework of general relativity in the form of a metric theory of acceleration and gravity, based on quasi-Riemann geometry.

These postulates, and the additional assumption that when gravitational fields are present nearby, space-time takes the form of a quasi-Riemannian manifold endowed with a metric curvature of the form $ds^2 = g_{ik} \, dx^i \, dx^j$, led Einstein to discover his famous gravitational field equations given below:

$$R_{ik} - (1/2) \, g_{ik} \, R = (8\pi G / c^2) \, T_{ik} \quad \text{(Einstein's Gravitational Field Equations)} \quad (5.1)$$

where, $g_{ik}$ is the metric tensor, $R_{ik}$ is the covariant Riemann curvature tensor. The left-hand side of the above equation is called the Einstein tensor or $G_{ik}$, which is the mathematical statement of space-time curvature that is reference frame independent and generally covariant. The right hand side $T_{\alpha\beta}$ is the stress-energy tensor which is the mathematical statement of the special relativistic treatment of mass-energy density, G is Newton's gravitational constant, and c the velocity of light.

For comparison purposes, we present the EMQG equations (reference 1) for the classical gravitational field where the gravitational field is not *too strong*, or *too weak*:

$$\nabla^2 \phi - (1/c^2) \, \partial^2 \phi / \partial t^2 = 4\pi G \, \rho(x,y,z,t) \quad (5.2)$$

where $\phi$ represents the classical Newtonian potential in absolute CA space and time units and $\rho(x,y,z,t)$ represents the **absolute** mass density distribution (that can be time varying) as measured from an observer at relative rest from the center of mass. This is a modified Poisson's equation, where the first term corresponds to the Poisson term, and the second term corresponds to the delay in the propagation of the graviton particles originating from the mass distribution. In EMQG, all distance units are expressed in absolute cellular automata space units in a 3D rectangular cell grid, and time as a count of the elapsed clock cycles. In other words, space is measured by counting the number of cells between two points (cells). Time is measured by counting the number of clock cycles that has elapsed between two events. The acceleration vector **a** for an average virtual particle at point (x,y,z) in CA space from the center of mass can be obtained from the gravitational potential $\phi$ at this point by the derivative of the potential as follows:

$$\mathbf{a} = \nabla \phi \qquad (5.3)$$

A detailed description of these equations are given in reference 1.

Einstein's law of gravitation (eq 11.1) cannot be arrived at by any 'rigorous' proof. The famous physicist S. Chandrasekhar writes (ref 37):

*"... It seems to this writer that in fact no such derivation exists and that, at the present time, no such can be given. ... It is the object of this paper to show how a mixture of physical reasonableness, mathematical simplicity, and aesthetic sensibility lead one, more or less uniquely, to Einstein's field equations."*

The principle of equivalence (in its strong form) is incorporated in the above framework by the assertion that all accelerations that are caused by either gravitational or inertial forces are **metrical** in nature. More precisely, the presence of acceleration caused by either an inertial force or a gravitational field modifies the geometry of space-time such that it is a quasi-Riemannian manifold endowed with a metric.

Furthermore, point particles move in gravitational fields along geodesic paths governed by the equation:

$$d^2x^i / ds^2 + \Gamma_{jk}^{\;i} (dx^j / ds)(dx^k / ds) = 0 \quad \text{... Equation for the geodesics} \qquad (5.4)$$

The most striking consequence of general relativity is the existence of curved 4D space-time specified by the metric tensor $g_{ik}$. In EMQG theory, the meaning of the geodesic is very simple; it is the path taken by light or matter through the falling virtual particles undergoing acceleration, in the absence of any other external forces. We will find later that curvature can be completely understood at the particle level. Furthermore, we will see that the principle of equivalence is a pure particle interaction process, and not a fundamental rule of nature. Before we can show this, we must carefully review the principle of equivalence from the context of general relativity theory.

## 6. THE PRINCIPLE OF EQUIVALENCE AND GENERAL RELATIVITY

*"I have never been able to understand this principle (principle of equivalence) ... I suggest that it be now buried with appropriate honors."*
- Synge: Relativity- The General Theory

It should be noted that Einstein did not explain the origin of inertia in general relativity. Instead he relied on the existing Newtonian theory of inertia. Inertia was described by Newton in his famous law: F=MA; which states that an object resists being accelerated. A force (F) is required to accelerate an object of mass (M) to an acceleration (A). Since acceleration is a form of motion, it would seem that a reference frame is required in order to gauge this motion. But this is not the case in Newtonian physics. All observers agree as to which frame is actually accelerating by finding out which frames has a force associated with it. Only non-accelerated frames are relative. Einstein did not elaborate on this anomaly, or provide a reason why the inertial and gravitation masses are equal. This still remains as a postulate in his theory. The principle of equivalence has been tested to great accuracy. The equivalence of inertia and gravitational mass has been verified to an accuracy of one part in about $10^{-15}$ (ref 24).

Einstein's general theory of relativity is considered a "classical" theory, because matter, space, and time are treated as continuous classical variables. It is known however, that matter is made of discrete particles, and that forces are caused by particle exchanges as described by quantum field theory. A more complete theory of gravity should encompass a detailed quantum process for gravity involving particle interactions only.

Inertia ought to be explained at the particle level as well, and should somehow be tied in to quantum gravity in a deep way according to the principle of equivalence. But, until recently there has been no adequate explanation for the origin of inertia. The next section summarizes some recent work on this problem, which has become the basis of EMQG.

## 7. THE PHYSICAL PROPERTIES OF THE MASSEON PARTICLE

In order to understand the principle of equivalence on the quantum level, we must postulate the existence of a new elementary particle. This particle is the most elementary form of matter or anti-matter, and carries the lowest possible quanta of low level gravitational 'mass charge'. This elementary particle is called the masseon particle (and also comes as anti-masseons, the corresponding anti-particle). The masseon is postulated to be the most elementary mass particle and readily combines with other masseons through a new, unknown hypothetical force coupling which we call the 'primal force'. Presumably, the primal force comes in positive and negative 'primal charge' types. The proposed mediator of this force is called the 'primon' particle. Since the masseon has not yet been detected, we can safely assume that the primal force is very strong. It is not necessary to understand the exact nature of the primal force to achieve the important results of EMQG. Suffice it to say that the primal force binds together masseon particles to make all the known fermion particles of the standard model. The masseon carries the lowest possible quanta of positive gravitational 'mass charge'. Low level gravitational 'mass charge' is defined as the (probability) fixed rate of emission of graviton particles in close analogy to electric charge in QED. Recall that the graviton is the vector boson of the pure gravitational force. Gravitational 'mass charge' is a fixed constant in EMQG, and is analogous to the fixed electrical charge concept. Gravitational 'mass charge' is **not** governed by the ordinary physical laws of *observable* mass, which appear as 'm' in the various physical theories. This includes Einstein's special relativity theory:

$E=mc^2$ or $m = m_0 (1 - v^2/c^2)^{-1/2}$ (15.41)

This is why we call it gravitational 'mass charge' or sometimes called the *low-level* mass of a particle, and this should not be confused with the ordinary observable inertial or observable gravitational mass. It will be assumed that when the low level mass is used in this paper, we are talking about the low level gravitational 'mass charge' property of a particle, and the associated graviton exchange process.

Masseons simultaneously carry a positive gravitational 'mass charge', and either a positive or negative electrical charge (defined exactly as in QED). Therefore, masseons also exchange photons with other masseon particles. It is important to note that the graviton exchange process is responsible for the low-level gravitational interaction only, which is not directly accessible to our measurements, and is also masked by the presence of the electromagnetic force component in all gravitational measurements. Masseons are fermions with half integer spin, which behave according to the rules of quantum field theory. Gravitons have a spin of one (not spin two, as is commonly thought), and travel at the speed of light. This paper addresses the gravitational and electromagnetic force interactions only, and the strong and weak nuclear forces are ignored here. Presumably, masseons also carry the strong and weak 'nuclear charge' as well.

Anti-masseons carry the lowest quanta of negative gravitational 'mass charge'. Anti-masseons also carry either positive or negative electrical charge, with electrical charge being defined according to QED. An anti-masseon is always created with ordinary masseon in a particle pair as required by quantum field theory (specifically, the Dirac equation). In EMQG, the anti-masseon is the negative energy solution of the Dirac equation for a fermion, where now the **mass** is taken to be 'negative' as well. Ordinarily, the standard model requires that the mass of any anti-particle is always positive, in order to comply with the principle of equivalence, or $M_i=M_g$. In EMQG, the principle of equivalence is not taken to be an absolute law of nature, and is definitely grossly violated for anti-particles. The anti-particles have positive inertia mass and *negative* gravitational mass, or $M_i=-M_g$.

Thus, a beautiful symmetry exists between EMQG and QED for gravitational and electromagnetic forces. The masseon-graviton interaction becomes almost identical to the electron-photon interaction. There are only two differences between these forces. First, the ratio of the strength of the electromagnetic over the gravitational forces is on the order of $10^{40}$. Secondly, there exists a difference in the nature of attraction and repulsion between positive and negative gravitational 'mass-charges' (as detailed in the table #1 and 2).

In QED, the quantum vacuum as a whole is electrically neutral because the virtual electron and positron (negative electron) particles are always created in particle pairs with equal numbers of positive and negative electrical charge. In EMQG, the quantum vacuum is also gravitationally neutral for the **same** reason. At any given instant of time, there is a 50-50 mixture of virtual gravitational 'mass charges', which are carried by the virtual masseon and anti-masseon pairs. These masseon pairs are created with equal and opposite gravitational 'mass charge'. This is the reason why the cosmological constant is zero (or very close to zero). Half the graviton exchanges between quantum vacuum particles result in attraction, while the other half result in repulsion. To see how this works, we will closely examine how masseons and anti-masseons interact.

The following tables summarize the fundamental electron and masseon force interactions:

**TABLE #1     EMQG MASSEON - ANTI-MASSEON GRAVITON EXCHANGE**

|  | (DESTINATION) | |
| --- | --- | --- |
| (SOURCE) | MASSEON | ANTI-MASSEON |
| MASSEON | attract | attract |
| ANTI-MASSEON | repel | repel |

**TABLE #2     QED ELECTRON - ANTI-ELECTRON PHOTON EXCHANGE**

|  | (DESTINATION) | |
| --- | --- | --- |
| (SOURCE) | ELECTRON | ANTI-ELECTRON |
| ELECTRON | repel | attract |
| ANTI-ELECTRON | attract | repel |

In QED, if the source particle is an electron, it emits photons whose wave function induces repulsion when absorbed by a destination electron, and induces attraction when absorbed by a destination anti-electron. Similarly, if the source is an anti-electron, it emits photons whose wave function induces attraction when absorbed by a destination electron, and induces repulsion when absorbed by a destination anti-electron.

In EMQG, if the source particle is a masseon, it emits gravitons whose wave function induces attraction when absorbed by a destination masseon, and induces attraction when absorbed by a destination anti-masseon. If the source is an anti-masseon, it emits gravitons whose wave function induces repulsion when absorbed by a destination masseon, and induces repulsion when absorbed by a destination anti-masseon. This subtle difference in the nature of graviton exchange process is responsible for some major differences in the way that low-level gravitational 'mass charge' and the electrical charges operate.

It is convenient to think of the photon as occurring in photon and anti-photon varieties (the photon is its own anti-particle). Similarly, the graviton comes in graviton and anti-graviton varieties. Thus, we can say that the masseons emit gravitons, and anti-masseons emit anti-gravitons. The absorption of a graviton by either a masseon or anti-masseon induces attraction. The absorption of an anti-graviton by either a masseon or anti-masseon induces repulsion. Similarly, we can say the electrons emit photons and anti-electrons emit anti-photons. The absorption of a photon by an electron induces repulsion, and the absorption of a photon by an anti-electron induces attraction. The absorption of an anti-photon by an electron induces attraction, and the absorption of an anti-photon by an anti-electron induces repulsion.

## 7.1 THE QUANTUM VACUUM AND VIRTUAL MASSEON PARTICLES

What virtual particles are present in the quantum vacuum? In QED, it is virtual electrons and anti-electrons (and virtual muons and tauons), along with the associated virtual photons. In the standard model of particle physics the quantum vacuum consists of all varieties of virtual fermion and virtual boson particles representing the known virtual matter and virtual force particles, respectively. This includes virtual electrons, virtual quarks, virtual neutrinos for fermions, and virtual photons, virtual gluons, and virtual W and Z bosons for the bosons. In EMQG, we restrict ourselves to the study of gravity and electromagnetism. Therefore, the EMQG quantum vacuum consists of the virtual masseons and virtual anti-masseons, and the associated virtual photons and virtual graviton particles (sometimes, virtual masseon combine to form virtual electrons, etc). Recall that ordinary matter consists only of real masseons bound together in certain combinations to form the familiar elementary particles. We now ask how the virtual electrons/positrons of the QED vacuum behave in the vicinity of a real electrical charge. We want to compare this with virtual masseon and virtual anti-masseon near a large real mass-charge like the earth in our EMQG formulation.

First, we review how the QED quantum vacuum is affected by the introduction of a real negative electrical charge. According to QED, the nearby virtual particle pairs become **polarized** around the central charge. This means that the virtual electrons of the quantum vacuum are repelled away from the central negative charge, while the virtual positrons are attracted towards the central negative charge. Thus for real electrons the vacuum polarization produces charge screening, which reduces the charge of a real electron, when measured over relatively long distances. According to QED, each electron is surrounded by a cloud of virtual particles that winks in and out of existence in pairs lasting a tiny fraction of a second, and this cloud is always present and acts like an electrical shield against the real charge of the electron. Recently, a team of physicists led by D. Koltick of Purdue University in Indiana reported that charge screening of an electron has been observed (ref. 33) experimentally at the KEK collider. They fired high-energy particles at electrons and found that the effect of this cloud of virtual particles was reduced the closer a high-energy particle penetrated towards the electron. They report that the effect of the higher charge for an electron that has been penetrated by particles accelerated to an energy 58 giga-electron volts, was equivalent experimentally to a fine structure constant of 1/129.6. This agreed well with their theoretical prediction of 1/128.5.

Next we study how the EMQG quantum vacuum is affected by the introduction of a large mass. According to EMQG, the quantum vacuum virtual masseon particle pairs are **not** polarized near a large mass, as we found for electrons (as can be seen from table #1 above). The virtual masseon and anti-masseon pairs are *both* attracted towards the mass. This *lack* of polarization which results is the main difference between electromagnetism and gravity. A large gravitational mass (like the earth) does **not** produce vacuum polarization of virtual particles. In gravitational fields, all the virtual masseon and virtual anti-masseon particles of the vacuum have a net average statistical acceleration directed towards a large mass, and produces a net inward (acceleration vectors only) flux of quantum vacuum virtual masseon/anti-masseon particles that can, and do affect other masses placed nearby. In contrast to this, an electrically charged object ***does*** produce vacuum polarization in QED; where the positive and negative electric charges accelerates towards and away, respectively from the charged object. Hence, there is no energy contribution to other electrical test charges placed nearby (from the vacuum particles only), because the charged vacuum particles contributes equal amounts of force contributions from both directions.

We will see that in gravitational fields like the earth, the ***lack*** of vacuum polarization is responsible for the weak equivalence principle. This is because the electrically charged quantum vacuum masseons/anti-masseon particles can act in unison against a test mass dropped on the earth. Had there been vacuum polarization for masseons, the vacuum particles would act in the two opposite directions, and hence no net vacuum action would result against a test mass. Now we are in a position to state the basic postulates of EMQG theory.

## 7.2  VIRTUAL MASSEON FIELD NEAR A SPHERICAL MASS IN EMQG

Our first application of EMQG theory is to determine the quantum nature of the gravitational field for a spherically symmetrical large mass. In general relativity, Einstein's field equation has been solved for this special case (ref. 39), and the solution is called the Schwarzschild metric and given by:

$$ds^2 = dr^2 / (1 - 2GM/(rc^2)) - c^2 dt^2 (1 - 2GM/(rc^2)) + r^2 d\Omega^2 \quad (7.21)$$

where $d\Omega^2 = d\theta^2 + \sin^2\theta\, d\phi^2$

This is a complete mathematical description of the space-time curvature (the metric in polar coordinates)) near the large spherical mass in spherical coordinates. This equation describes the path that light or matter takes through curved 4 D Minkowski space-time. We will find that this solution is a very good *approximation* to the gravitational field. There are, however, hidden quantum processes involving the virtual particles in EMQG that are responsible for this curvature, and for the very tiny inaccuracy of this metric due to a slight violation of the principle of equivalence.

A large spherical mass turns out to be an excellent example for EMQG, because it has a very simple motion associated with the virtual particles that make up the surrounding quantum vacuum (in absolute CA units). The normal background motion of virtual particle creation and annihilation process near a massive spherical distribution of matter is distorted when compared to the vacuum in empty space far removed from any matter. Surrounding a large spherically symmetrical mass like the earth, the virtual particles created in the quantum vacuum have a net average acceleration vector that is directed downward towards the earth's center along radius vectors. (Note: We are ignoring the mutual interactions of the vacuum particles, which are why this statement is statistical in nature.) The cause of this downward acceleration of the vacuum is graviton exchanges between the earth and the virtual masseon particles of the quantum vacuum (postulate #2), which propagate at the speed of light (in absolute units). At any one instant, the vacuum particles have random velocity vectors which point in all directions, even including the *up* direction. However, the acceleration vectors are generally coordinated in the downward direction. The closer the virtual particles are to the earth, the greater the acceleration, as you would expect from the inverse square law of graviton exchanges. The average net statistical acceleration of this stream of virtual particles is directed downward, and varies with the height 'r'. This accelerated vacuum 'stream' plays the most important role in the dynamics of gravity in EMQG, and also naturally ties in with the problem of inertia and the equivalence principle. In fact, we will see that the average net acceleration vector of the vacuum particles at each point in space surrounding the earth, at its interaction with test masses and light is equivalent to the Schwarzschild 4D space-time metric given above. This is because the average net acceleration vector of the charged virtual particles at each point in space surrounding the mass guides the motion of the electrically charged free masseon particles or photons through the electromagnetic force. We will show this mathematically in section 9.6.

The magnitude and direction of the net average statistical acceleration of the virtual particles at point r above the earth (the direction is along the radius vectors) can be easily found from Newton's inverse square law of gravitation ($a = GM/r^2$). It is also possible to calculate this from the basic EMQG equations for a general, slow mass distribution (equation 17.22 and 17.23 of ref. 1). A complex calculation of the state of motion of the vacuum particles is not required in the case for large spherically symmetrical masses like the earth, because of the simplicity of the virtual particle motion.

When a small test mass moves through the space surrounding the earth, the electromagnetic interactions between the real charged masseon particles in the mass with respect to the virtual charged particles quantum vacuum dominates over the pure graviton exchange process between the mass and the earth. This electromagnetic component plays the **major** role in the dynamics of motion of a nearby test mass. From postulate #2, the real masseon particles consisting of the earth exchanges gravitons with the virtual masseons of the quantum vacuum, causing a downward acceleration of the quantum vacuum of 1g. If we now introduce a test mass near the earth, according to postulate #2 all the real masseons making up the test mass will fall at the same average rate as that of the net statistical average of virtual particles of the vacuum. This is due to the relatively strong electromagnetic force acting between the electrically charged virtual masseons of the vacuum and the real masseons of the test mass.

Therefore, based on the Newtonian principles of how ordinary matter falls, the net average acceleration 'a' of a virtual particle in the vacuum with respect to height of the test mass, along the radius vector '**r**' towards the center of the earth is given by:

The net statistical average acceleration vector:   $\mathbf{a} = GM / \mathbf{r}^2$      (7.22)

where **r** is the distance vector along the radius from the center of the earth to a typical virtual particle, G is the Newtonian Universal Gravitation constant, and M is the mass of the earth.

**Note:** We have not proved that equation (7.22) is correct. Instead, it is based on the observation of the motion of a test mass near the earth. However, this equation can be derived from the semi-classical EMQG equations of motion.

To fully account for the gravitational field around a spherically symmetrical massive object, the motion of light near the object must also be accounted for. We will find that the altered behavior of light near a massive object drastically modifies the nature of equation (7.22). This equation is based on absolute cellular automata 3D space and time units. Relativistic curved 4D space-time is an emergent phenomena from this process, because of the way that light and matter behaves in this 'accelerated stream of virtual particles' near the earth. This alters the nature of equation 7.22, which now has to be specified in absolute CA units. This is because the acceleration (a=dv/dt, and velocity v=dx/dt) involves distance and time measurements.

## 8. THE PRINCIPLE OF EQUIVALENCE

*"The principle of equivalence performed the essential office of midwife at the birth of general relativity, but, as Einstein remarked, the infant would never have got beyond its long clothes had it not been for Minkowski's concept [of space-time geometry]. I suggest that the midwife be now buried with appropriate honors and the facts of absolute space-time faced."*                    *- Synge*

The principle of equivalence means different things to different people, and to some it means nothing at all as can be seen in the quotation above. The equality of inertial and gravitational mass is only known to be true strictly through observation and experience. Is this equivalence exact, though? Since the principle of equivalence cannot be currently traced to deeper physics, we can never say that these two mass types are *exactly* equal. Currently, we can only specify the accuracy to which the two mass types have been shown *experimentally* to be equal.

How is the principle of equivalence defined? Well, there are two main formulations of the principle of equivalence. The strong equivalence principle states that the results of any given physical experiment will be precisely identical for an accelerated observer in free space as it is for a non-accelerated observer in a perfectly uniform gravitational field. A weaker form of this postulate restricts itself to the laws of motion of masses only. In other words, the laws of motion of identical masses on the earth are identical to the same situation inside an accelerated rocket (at 1g). Technically, this holds only at a point near the earth. It can be stated that objects of the different mass fall at the same rate of acceleration in a uniform gravity field. In regards to the strong equivalence principle, Synge writes:

*"... I never been able to understand this Principle ... Does it mean that the effects of a gravitational field are indistinguishable from the effects of an observer's acceleration? If so, it is false. In Einstein's theory, either there is a gravitational field or there's none, according as the Riemann tensor does not or does vanish. This is an absolute property. It has nothing to do with any observer's world line ... The principle of equivalence performed the essential office of midwife at the birth of general relativity, but, as Einstein remarked, the infant would never have got beyond its long clothes had it not been for Minkowski's concept [of space-time geometry]. I suggest that the midwife be now buried with appropriate honors and the facts of absolute space-time faced."*

Few physicists would doubt the validity of his statement. Synge has hit on an important point in regards to the nature of the equivalence principle and space-time. He is right to say that "*either there is a gravitational field or there's none, according as the Riemann tensor does not or does vanish. This is an absolute property* (of space near masses)". What he means is that the Riemann tensor describing curvature is there, or is not there, depending on whether or not there is a large mass present to distort space-time. (in other words, whether there exists a global space-time curvature or not). The existence of a **global** space-time curvature reveals whether you are in a gravitational field. In an accelerated frame, the space-time curvature is local to your motion only, and is not global property of space-time.

According to EMQG, if a large mass is present, the mass emits huge numbers of graviton particles, and distorts the surrounding virtual particles of the quantum vacuum. In an accelerated frame, there are very few gravitons, and the quantum vacuum is not affected.

However, an observer in the accelerated frame 'sees' the quantum vacuum accelerating with respect to his frame, and hence the space-time distortion. However, the quantum vacuum *still remains undisturbed*. Thus in EMQG, the equivalence principle is regarded as being a coincidence due to quantum vacuum appearing the same for accelerated observers and for observers in gravitational fields.

Recently, some theoretical evidence has appeared to suggest that the strong equivalence principle does *not* hold in general. First, if gravitons could be detected experimentally with a new and sensitive graviton detector (which is not likely to be possible in the near future), we would be able to distinguish between an inertial frame and a gravitational frame with this detector. This is possible because inertial frames would have virtually no graviton particles present, whereas the gravitational fields like the earth have enormous numbers of graviton particles. Thus, we have performed a physics experiment that can detect whether you are in a gravitational field or an accelerated frame. Secondly, recent theoretical considerations of the emission of electromagnetic waves from a uniformly accelerated charge, and the lack of radiation from the same charge subjected to a static gravitational field leads us to the conclusion that the strong equivalence principle does not hold for radiating charged particles. Stephen Parrott (ref 23) has done an extensive analysis of the electromagnetic energy released from an accelerated charge in Minkowski space and a stationary charge in Schwarzchild space. He writes in his paper on "Radiation from a Uniformly Accelerated Charge and the Equivalence Principle":

*"It is generally accepted that any accelerated charge in Minkowski space radiates energy. It is also accepted that a stationary charge in a static gravitational field does not radiate energy. It would seem that these two facts imply that some forms of Einstein's Equivalence Principle do not apply to charged particles.*

*To put the matter in an easily visualized physical framework, imagine that the acceleration of a charged particle in Minkowski space is produced by a tiny rocket engine attached to the particle. Since the particle is radiating energy, that can be detected and used, conservation of energy suggests that the radiated energy must be furnished by the rocket. We must burn more fuel to produce a given accelerated world line than we would to produce the same world line for a neutral particle of the same mass. Now consider a stationary charge in Schwarzchild space-time, and suppose a rocket holds it stationary relative to the coordinate frame (accelerating with respect to local inertial frames). In this case, since no radiation is produced, the rocket should use the same amount of fuel as would be required to hold stationary a similar neutral particle. This gives an experimental test by which we can determine locally whether we are accelerating in Minkowski space or stationary in a gravitational field - simply observe the rocket's fuel consumption."*

He does a detailed analysis of the energy in Minkowski and Schwarzchild space-time, and shows that strong principle of equivalence does not hold for charged particles.

As for the weak equivalence principle, we can now only specify the accuracy as to which the two different mass types have been shown *experimentally* to be equal in an inertial and gravitational field. In EMQG, we show that the equivalence principle follows from lower level physical processes, and the basic postulates of EMQG. We will see that mass equivalence arises from the equivalence of the force generated between the net statistical average acceleration vectors of the matter particles inside a mass interacting with the surrounding quantum vacuum virtual particles inside an accelerating rocket. The *same* force occurs between the matter particles and virtual particles for a mass near the earth. We will find that equivalence is *not* perfect, and breaks down when the accuracy of the measurement approaches $10^{-40}$!

Basically, the equivalence principle arises from the *reversal* of the net statistical average acceleration vectors between the charged matter particles and virtual charged particles in the famous Einstein rocket, with the same matter particles and virtual particles near the earth. To fully understand the hidden quantum processes in the principle of equivalence on the earth, we will detail the behavior of test masses and the propagation of light near the earth. Equivalence is shown to hold for both stationary test masses and for free-falling test masses.

First we derive the principle of equivalence for the motion of ordinary masses. Next, we show that the quantum principle of equivalence holds for elementary particles. Next, we will demonstrate that equivalence also holds for large spherical masses with considerable self-gravity (and self-energy) such as the earth with a hot molten core, and the moon with a considerably colder core, with respect to a third mass like the sun. We will see that if both the earth and the moon fall towards the sun, they would arrive at the same time to a high degree of precision in the framework of EMQG. Finally, we examine the principle of equivalence and curved Minkowski 4D space-time curvature.

8.1     MASSES INSIDE AN ACCELERATED ROCKET AT 1g

In figure #1, there are two different masses at rest on the floor of a rocket which is accelerated upwards at 1 g far from any gravitational sources. The floor of the rocket experiences a force under the mass '2M' that is twice as great as for the mass 'M'. In Newtonian physics, the inertial mass is defined in precisely this way, the force 'F' that occurs when a mass 'M' is accelerated at rate 'g' as given by F=Mg. The quantum inertia explanation for this is that the two masses are accelerated with respect to the net average statistical motion of the virtual particles of the vacuum by the rocket. Since mass '2M' has twice the masseon particle count as mass 'M', the sum of all the tiny electromagnetic forces between the virtual vacuum and the masseon particles of mass '2M' is twice as great as compared to mass 'M', i.e. for mass 'M', $F_1$=Mg and for mass '2M', $F_2$=2Mg=2$F_1$. Because the particles that make up the masses do not maintain a net zero acceleration with respect to the virtual particles, a force is always present from the rocket floor (figure 1).

In figure #2, the two different masses (M and 2M) have just been released and are in free fall inside the rocket. According to Newtonian physics, no forces are present on the two masses since the acceleration of both masses is zero (the masses are no longer attached to the rocket frame). The two masses hit the rocket floor at the same time. The quantum inertia explanation for this is trivial. The net acceleration between all the real masseons that make up both masses and virtual masseon particles of the vacuum is a net (statistical average) value of zero. The rocket floor reaches the two masses at the same time, and thus unequal masses fall at the same rate inside an accelerated rocket.

8.2   MASSES INSIDE A GRAVITATIONAL FIELD (THE EARTH)

In figure #3, there are the same two masses (2M and M) which are at rest on the surface of the earth. The surface of the earth experiences a force under mass '2M' that is twice as great as for that under mass 'M'. The reason for this is that the two stationary masses do not maintain a net acceleration of zero with respect to the net statistical average acceleration of the virtual masseons in the neighborhood. This is because the virtual particles are all accelerating towards the center of the earth ($\mathbf{a}=GM/\mathbf{r}^2$) due to the graviton exchanges between the real masseons consisting of the earth and the virtual masseons of the vacuum. Since mass '2M' has twice the masseon particles as mass 'M', the sum of all the tiny electromagnetic forces between the virtual masseon particles of the vacuum and the real masseon particles of mass '2M' is twice as great as that for mass 'M'. Thus, a force is required from the surface of the earth to maintain these masses at rest, mass '2M' having twice the force of mass 'M'. The physics of this force is the same as for figure #1 in the rocket, but with the acceleration frames of the virtual charged masseons and the real charged masseon particles of the mass being reversed (with the exception of the direct graviton induced forces on the masses, which is negligible). Equivalence between the inertial mass 'M' on a rocket moving with acceleration 'A', and gravitational mass 'M' under the influence of a gravitational field with acceleration 'A' can be seen to follow from Newton's laws as follows:

$F_i = M(A)$      ...inertial force opposes the acceleration A of the mass 'M' in rocket.
$F_g = M(GM_e/r^2)$ ...gravitational force where **$GM_e/r^2$ is now virtual particle acceleration**.

Under gravity, the magnitude of the gravitational field acceleration is $A=GM_e/r^2$, which is the same as the magnitude of the acceleration of the rocket. From the reference frame of an average accelerated virtual particle on earth, a virtual masseon particle 'sees' the real masseon particles of the stationary mass M accelerating in exactly the same way as an average stationary virtual masseon in the rocket 'sees' the accelerated mass particles in the rocket. In other words, the vacuum state appears the same from both of these reference frames. We have illustrated equivalence in a special case; between an accelerated mass M and a stationary gravitational mass 'M'. Equivalence holds because $GM_e/r^2$ represents the net statistical average downward acceleration vector of the virtual masseons with respect to the earth's center, and is **equal** to the acceleration of the rocket. Newton's law of gravity was rearranged here to emphasize the form F=MA for *gravitational mass* so that we can see that the **same** electromagnetic force summation process for real masseons of the mass occurs under gravity as it does for accelerated mass. Thus the same processes at work in inertia are also present in gravitation.

This example shows why both the masses of figure 1 are equivalent to the masses in figure 3. The force magnitude is the same because the calculation of the force involves the same sum of all the tiny electromagnetic forces between the virtual charged masseon particles and the real masseon particles of the mass. The only difference in the physics of the masses in figure 1 is that the relative motions of all the tiny electromagnetic force vectors are reversed. The other difference is that large numbers of graviton particles (that originate from the earth's mass) slightly unbalances perfect equivalence between the masses falling on the earth. The larger mass has the largest graviton flux.

**Note:** There is a *very small* discrepancy in the equivalence principle for unequal masses in free fall near the earth which is caused by the excess graviton exchange force for the heavier mass. This discrepancy in the free fall rate of test masses near the earth is extremely minute in magnitude because there is a ratio of about $10^{40}$ in the field strength existing between the electromagnetic and gravitational forces. In principle it could be measured by extremely sensitive experiments, if two test masses are chosen with a very large mass difference.

In figure #4, two different masses are in free fall near the surface of the earth, and no external forces are present on the two masses. The two masses hit the earth at the same time. The net statistical average acceleration of the real masseon particles that make up the masses and virtual charged masseon particles of the vacuum is still zero, because this process is dominated by the electromagnetic force (the direct graviton exchanges are negligible). The electromagnetic forces between the virtual particles and the matter particles of the test mass dominates the interactions, because the electromagnetic force is $10^{40}$ times stronger than the graviton component. Although mass '2M' has twice the gravitational force due to twice the number of graviton exchanges, this is totally swamped out by the electromagnetic interaction, and the accelerated virtual particles and the test masses are in a state of electromagnetic equilibrium as far as acceleration vectors are concerned. Both masses fall at the same rate (neglecting the slight imbalance of the note above).

### 8.3  MICROSCOPIC EQUIVALENCE PRINCIPLE OF PARTICLES

Does the weak equivalence principle hold for an elementary particle? For example, does a neutron and an electron simultaneously dropped on the surface of the earth fall at the same rate (ignoring stray electrical charge effects)? Is this equivalent to the same experiment performed inside a rocket that is accelerating at 1 g? The answer to all these questions is yes. In fact, the equivalence principle has actually been experimentally verified for the case of a neutron in a gravitational field (ref. 40).

An astute observer may have questioned why **all** the virtual particles (virtual neutrons, virtual electrons, virtual quarks, etc, all consisting of different masses) are accelerating downwards towards the earth with the same acceleration in our EMQG model. Certainly inside an accelerated rocket an observer stationed on the floor will view *all* the virtual particles of the quantum accelerating with respect to him at the same rate, no matter what the masses of the virtual particles are. This is simply because the floor of the rocket moves upward at 1g, giving the *illusion* (to an observer on the floor of the rocket) that virtual particles of different mass are accelerating at the same rate. Since the masses of the different types of virtual particles are **all different** according to the standard model of particle physics, why are they all falling at the same rate on the earth? Here, the cause of the acceleration is graviton exchanges with the earth. Since we are trying to derive the equivalence principle from fundamental concepts, we cannot invoke this principle to state that the virtual particles must be accelerating at the same rate.

We can trace why all quantum vacuum virtual particles are accelerating at the same rate on the earth to the existence of the virtual masseon particle. All particles with mass (virtual or not) are composed of combinations of the fundamental "masseon" particle, which carries just one fixed quanta of mass (postulate #2). Since all virtual masseon particles exchange the same fixed flux of gravitons with the earth, the virtual masseons are all accelerated at the same rate. However, masseons can bind together to form the familiar particles of the standard model such as virtual electrons, virtual positrons, virtual quarks, etc. or even unknown species of virtual particles. According to postulate #2, masseons carry both gravitational 'mass charge' and ordinary electrical charge. However, the electromagnetic interactions (photon exchanges) will work to equalize the fall rate (from the point of view of acceleration vectors) of virtual masseons that momentarily combine to make virtual particles like virtual electrons and virtual quarks. If a virtual quark consists of say 100 bound masseons (the actual number is not known), the graviton exchanges would normally be cumulative, and 100 times more acceleration will be imparted to the virtual quark than a single virtual masseon. However, virtual masseons dominate the quantum vacuum since they are the fundamental mass particle, and do not have to bond with other masseons to exist. Therefore the lone, unbound virtual masseon is by far the most common virtual mass particle in the quantum vacuum (this is illustrated in figures 5 and 6).

No matter how many virtual masseons combine to give other virtual particles, the local electromagnetic interaction between the far more numerous virtual masseons and the virtual quark (or any other virtual particle) will equalize the fall rate. This process works like a *microscopic version* of the EMQG weak principle of equivalence, for falling virtual particles, with the same action occurring on the particle level as what happens for large falling masses discussed in section 8.3. Figures 5 and 6 shows the microscopic equivalence principle at work for a free falling mass in an accelerated frame and in a gravitational field. To summarize, the electromagnetic forces from the free virtual masseons of the quantum vacuum (all falling at the same rate), dominates over the more familiar virtual particles that consist of combinations of masseons (like the virtual neutrons, electrons, quarks, and all other virtual particles). The virtual quark would normally fall faster than the virtual electron and the virtual electron faster than an individual virtual masseon. This is because many virtual masseons bound together exchange many more gravitons with the earth. However, the electromagnetic interaction between the far numerous virtual masseons of the vacuum, and the virtual masseons combined inside the virtual neutrons, quarks, and virtual electrons acts to equalize the fall rate, causing all virtual particles to fall at the same rate. Since the quantum vacuum background appears the same from the perspective of a mass on the surface of the earth, as for the same mass inside an accelerated rocket equivalence still holds.

## 8.4 EQUIVALENCE PRINCIPLE FOR THE SUN-EARTH-MOON SYSTEM

Will the weak equivalence principle hold for the following imaginary scenario, where the earth and the moon are simultaneously in free fall towards the sun with an acceleration of gravity '$G_{sun}$' of the sun? In other words, would the earth and moon arrive at the same time on the surface of the sun? Would they also arrive on the floor at the same time when free falling inside a huge rocket undergoing acceleration $G_{sun}$ in space (far from any other large masses)? This question is at the heart of the so-called metric theories of gravity. In metric theories such as general relativity, all objects with any kind of internal composition follow the natural curvature of space-time. This includes objects with considerable internal energy sources and self-gravity. Any deviation from perfect equivalence would constitute what is called the Nordtvedt effect (ref. 41), after the discoverer in 1968. Because the earth's core is molten and very hot, the earth contains a significant internal energy. The earth is also a large source of gravitational energy as well, which significantly distorts the nearby virtual particles of the quantum vacuum. Contrast this with the moon, which is relatively cool and less energetic, with considerably less gravitational energy.

To see if the weak equivalence applies here, we start with the situation where the earth and moon are 'dropped' simultaneously from a height 'h' inside a huge rocket (with negligible mass) accelerating with the same $G_{sun}$ as exists on the surface of the sun (figure 7). The result of this experiment is obvious. They both arrive on the floor of the rocket at the same time. Actually, it is the floor of the rocket that accelerates upwards and meets both bodies at the same time! However, now the virtual particles of the quantum vacuum are disturbed near these large bodies (by graviton exchanges), particularly the acceleration vectors of the virtual particles in close proximity with the earth and the moon (now tending to point towards the centers of the two bodies in these regions). This fact, however, does not affect the results of this experiment. The results are no different than if two small masses are dropped inside the rocket; again because the floor moves up to meet them at the same time. However, one must note the virtual particle pattern of figure 8.

The situation near the surface of the sun, where the earth and moon are 'dropped' from the same height 'h', is far more complex than inside the rocket. Now all three bodies disturb the virtual particles of the quantum vacuum! The sun sets up a strong $GM_s/r^2$ acceleration field consisting of virtual particles, which applies over long distances and points towards the center of the sun. The earth and moon also produce their own fields in their vicinity, although much weaker (figure 7). The sun's acceleration dominates over the surrounding space, except near the moon and near the earth; where some of the virtual particles actually are moving away from the sun (as happens near the surface of the night side of the earth). How can equivalence possibly hold in this scenario? Recall that we stated that it is the electromagnetic action of the virtual particles of the vacuum on the real particles inside the bodies that determines the motion of the earth and the moon undergoing gravitational acceleration. However, in different regions of the earth, the virtual particles are accelerating in different directions! Part of the answer to this problem is an important property exhibited by the graviton particle (postulate #2): the **principle of superposition**. This is a property also shared by the photon particle. The action of the gravitons originating from all three sources on a given virtual particle of the quantum vacuum yields a net acceleration that is the net vector sum of the action of all the gravitons received by the virtual particle.

To explain equivalence, we must first recall that equivalence only holds in a sufficiently small region of space (technically, at a given point above the sun) when compared to the equivalent accelerated reference frame. This is because the acceleration of the sun varies with the distance 'r' from the sun's center, whereas inside an accelerated rocket it does not vary with height. Secondly, we must recall that the motion of the virtual particles in the rocket is also disturbed near the vicinity of the earth and the moon. In fact, inside the rocket the virtual particles are also directed along the radius vectors of both the earth and the moon in their vicinity (figure 7). Yet, the sun and the moon still reach the floor of the rocket at the same time. Therefore, the quantum vacuum can be disturbed in the case of the free fall of the earth and the moon towards the sun, provided the total virtual particle pattern can be shown to match the case of the accelerated rocket.

A close study of figures 7 and 8 reveals that the quantum vacuum pattern *is the same* when viewed in the *correct* reference frame. In figure 7, the observer is stationed on the surface of the sun, so that we can see the reason why both bodies are attracted to the sun. Recall that the electromagnetic interaction between the acceleration vectors of the falling virtual masseons and the real masseons in the earth and moon is the primary reason for the attraction. The direct graviton action is negligible in comparison. Now in order to compare the two experiments of figure 7 and 8, the reference frame for the sun experiment should be equivalent to the rocket. In the rocket experiment, the frame chosen for our observer is outside the rocket (the quantum vacuum has a relative acceleration of zero) in order to understand the results. Therefore for the sun, the observer's frame should be in free fall, thus restoring the relative acceleration of the quantum vacuum to zero just as for the observer outside the rocket. When this is done, we have to correct the acceleration vectors of the virtual particles near the earth and moon in figure 7. It is easy to show that the result of this operation gives an identical result as figure 8. Therefore, equivalence holds in both experiments.

## 8.5 LIGHT MOTION IN A ROCKET: SPACE-TIME EFFECTS

We will examine three scenarios for the motion of light in a rocket which is accelerated upwards at 1 g (far from any gravitational sources). First, we study light moving from the floor of the rocket to the ceiling where it is detected by an observer. Next we look at light moving from the ceiling of the rocket to the floor where it is detected by an observer. Finally, we examine light moving parallel with the floor of the rocket, where it follows a curved path (figure 10).

### (A) LIGHT MOVING FROM THE FLOOR TO THE CEILING OF THE ROCKET

Here the light is positioned on the floor of the rocket which is being accelerating upwards at 1 g, and propagates in a straight line up to the observer on the ceiling. Meanwhile, the rocket has accelerated upwards while the light is in flight. What happens to the light? According to general relativity, an observer outside the rocket examines the light moving upward at the speed of light in a straight path. Meanwhile, according to general relativity, an observer inside the rocket stationed on the ceiling also observes the light moving upwards in a straight line towards him. He also observes that the light is red-shifted. He makes a measurement of the light velocity of the incoming red-shifted light with his measuring instruments (which were calibrated within his reference frame). He observes that the velocity of the red-light light is the same on the ceiling as he found when he previously checked his internal light sources with his calibrated instruments. In other words, the speed of light does not vary under all these circumstances. Closer examination reveals that the clocks in the ceiling differ from the clocks stationed on the floor. In particular, the clock on the floor of the rocket runs slower than one on that on the ceiling. Distances measurements are also affected. General relativity explains all these observations with the 4D space-time curvature existing inside accelerated frames. We will return to this example with our EMQG interpretation of these measurements.

### (B) LIGHT MOVING FROM THE CEILING TO THE FLOOR OF THE ROCKET

Here the light is positioned on the ceiling of the rocket which is accelerating upwards at 1 g, and propagates in a straight line down to the observer on the floor. Meanwhile, the rocket has accelerated upwards while the light is in flight. What happens to the light? According to general relativity, an observer outside the rocket observes the light moving downwards at the speed of light in a straight path. Meanwhile, according to general relativity, an observer inside the rocket stationed on the floor also observes the light moving downwards in a straight line towards him. He also observes that the light is blue-shifted. He makes a measurement of the light velocity of the incoming blue-shifted light with his measuring instruments (which were calibrated within his reference frame). He observes that the velocity of the blue-light light is the same on the floor as he found when he previously checked his internal light sources and with his calibrated instruments. In other words, the speed of light does not vary under all these circumstances. Closer examination reveals that the clocks in the floor differ from the clocks stationed on the ceiling. In particular, the clock on the ceiling of the rocket runs faster than one on that on the floor. Distances measurements are also affected. Again, general relativity explains all these observations with the 4D space-time curvature existing inside accelerated frames. We will return to this example with our EMQG interpretation of these measurements.

(C) LIGHT MOVING PARALLEL TO THE FLOOR OF THE ROCKET

Here the light leaves the light source on the left wall of the rocket which is accelerating upwards at 1 g, and propagates in a straight line towards the observer on the right wall (figure 10). Meanwhile, the rocket has accelerated upwards while the light is in flight. Therefore an observer in the rocket observes a curved light path. An observer outside the rocket sees a straight light path. According to general relativity, the space-time inside the rocket is curved (in the direction of motion), and light moves along the natural geodesics of curved 4D space-time. Meanwhile, the observer outside the rocket lives in flat-space time, and therefore observes light moving in a perfect straight line, which is the geodesic path in flat 4D space-time. We will return to this example with our EMQG interpretation of these measurements. Next we will examine all three scenarios on the surface of the earth.

8.6    LIGHT MOTION NEAR EARTH'S SURFACE - SPACE-TIME EFFECTS

We will examine the same three scenarios for the motion of light on the surface of the earth (1g), which is the same as for the rocket (1g) according to the principle of equivalence. We will ignore the variation of acceleration with height found on the earth, as well as the slight change in the direction of acceleration caused by acceleration vectors being directed along radius vectors. First, light moves from the floor of the room on the surface of the earth to the ceiling, where it is detected. Next, light is moving from the ceiling of the room to the surface of the earth where it is detected by an observer. Finally, light is moving parallel with the earth's surface from the left side of the room to the right, and follows a curved path (figure 9).

(A) LIGHT MOVING FROM THE FLOOR TO THE CEILING ON EARTH

Light is positioned on the floor of a room on the surface of the earth, and propagates in a straight line up to the observer on the ceiling. What happens to the light? According to general relativity, an observer outside the room in free fall observes the light moving upward at the speed of light in a straight path. Meanwhile, according to general relativity, an observer inside the room stationed on the ceiling also observes the light moving upwards in a straight line. He also observes that the light is red-shifted. He makes a measurement of the light velocity with his measuring instruments (which were calibrated within his reference frame) and observes that the velocity of light is the same on the floor as he found when he measured received light speed in his internal reference frame with the same instruments. In other words, the speed of light does not vary in all cases. Closer examination reveals that clocks measured in his reference frame differ from the clocks on the floor. In particular, the clock on the floor of the room runs slower than the one on the ceiling. Distances are also affected. In general relativity, all these conclusions follow directly from 4D space-time curvature.

(B) LIGHT MOVING FROM THE CEILING TO THE FLOOR ON EARTH

Here the light is positioned on the ceiling of the room on the surface of the earth, and propagates straight down to the observer on the floor. What happens to the light? According to general relativity, an observer outside the room in free fall observes the light moving downward at the speed of light in a straight path. Meanwhile, according to general relativity, an observer inside the room stationed on the floor also observes the light moving downwards in a straight line. He also observes that the light is blue-shifted. He makes a measurement of the light velocity with his measuring instruments (which were calibrated within his reference frame) and observes that the velocity of light is the same on the floor as he found when he measured received light speed in his internal reference frame with the same instruments. In other words, the speed of light does not vary in all cases. Closer examination reveals that clocks measured in his reference frame differ from the clocks on the ceiling. In particular, the clock on the ceiling of the room runs faster than the one on the floor. Distances are also affected. In general relativity, all these conclusions follow directly from 4D space-time curvature.

(C) LIGHT MOVING PARALLEL TO THE SURFACE OF THE EARTH

Here the light leaves the light source on the left wall of the room on the earth and propagates in a curved path towards the observer on the right wall (figure 9). Meanwhile, an observer in free fall towards the earth's surface sees light moving in a straight path. According to general relativity, and light moves along the natural geodesics of curved 4D space-time in the room. Meanwhile, an observer in free fall lives in flat 4D space-time, and hence an observer sees straight-line paths for light. In general relativity, all these conclusions are identical as for the observer accelerated in the rocket at 1g in accordance with the principle of equivalence. Now we look at 4D space-time curvature from the perspective of EMQG.

### 9. EQUIVALENCE PRINCIPLE AND SPACE-TIME CURVATURE

*"The relativistic treatment of gravitation creates serious difficulties. I consider it probable that the principle of the constancy of the velocity of light in its customary version holds only for spaces with constant gravitational potential."*

                                                          - **Albert Einstein  (in a letter to his friend Laub, August 10, 1911)**

In this section, we contrast the two different approaches to the problem of space-time curvature, and the propagation of light in a gravitational field: Einstein's General Relativity and EMQG theory. First we will derive the gravitational time dilation equation using the solution to Einstein's gravitational field equations for a spherical mass called the Schwarzschild metric. Next, we fully develop the EQMG theory of space-time curvature. From this, we calculate the quantity of space-time curvature using EMQG theory, and show that the results are the same.

9.1   GENERAL RELATIVISTIC 4D SPACE-TIME CURVATURE

General relativity accounts for the motion of light under all scenarios for a large spherical mass. General Relativity postulates space-time curvature in order to preserve the constancy of the light velocity in an accelerated frame or in a gravitational field. The solution of Einstein's gravitational field equation for the case of spherical mass distribution is given by the Schwarzchild metric (ref. 39):

$$ds^2 = dr^2 / (1 - 2GM/(rc^2)) - c^2 dt^2 (1 - 2GM/(rc^2)) + r^2 d\Omega^2 \qquad (9.11)$$

where $d\Omega^2 = d\theta^2 + \sin^2\theta \, d\phi^2$

This is a complete mathematical description of the space-time curvature near the large spherical mass in spherical coordinates in differential form called the 4D space-time metric. From this, it is easy to show (ref. 39) that the comparison of time measurements between a clock outside a gravitational field (called proper time $t(\infty)$) to a clock at distance r from the center of a spherical mass distribution (called the coordinate time $t(r)$) is given by:

$$t(r) = (1 - 2GM/(rc^2))^{-1/2} \, t(\infty) \qquad (9.12)$$

which follows from Schwarzchild metric directly.

Using the relationship $(1 - x)^{-1/2} \approx 1 - x/2$ when $x \ll 1$, and realizing the quantity $2GM/rc^2$ is very small (for the earth this is $\approx 10^{-9}$) we can write this as:

$$t(r) \approx (1 - GM/(rc^2)) \, t(\infty) \qquad (9.13)$$

This gives the amount of time dilation between a clock on the earth "$t(r)$" compared to a clock positioned at infinity "$t(\infty)$". From this, we see that clocks on the earth run slower then at infinity.

Similarly, from the metric, we find that the distance at point s( r ) follows as:

$$s(\infty) = (1 - 2GM/(rc^2))^{-1/2} \, s(r) , \qquad (9.14)$$

or we can also write this as:

$$s(r) \approx \left(1 - GM/(rc^2)\right)^{-1} s(\infty) \qquad (9.15)$$

This gives the amount of space distortion for rulers on the earth "$s(r)$" compared to rulers positioned at infinity "$s(\infty)$".

## 9.2 EMQG AND 4D SPACE-TIME CURVATURE

In order to understand space-time curvature and the principle of equivalence in regards to the equivalence of all light motion in an accelerated rocket compared with that on the surface of the earth, we must examine the effects of the background virtual particles on the propagation of light. The big question to consider here is this:

**Does the general downward acceleration of the virtual particles of the quantum vacuum near a large mass affect the motion of nearby photons? Or is the deflection of photons truly the result of an actual space-time geometric curvature (which holds down to the tiniest of distance scales), as required by the constancy of the light velocity in Einstein's special relativistic postulate**?

The answer to this important question hinges on whether our universe is truly a curved, geometric Minkowski 4D space-time on the smallest of distance scales, or whether curved 4D space-time results merely from the activities of quantum particles interacting with other quantum particles. EMQG takes the second view! According to postulate 4 of EMQG theory, light takes on the same general acceleration as the net statistical average value of quantum vacuum virtual particles, through a 'Fizeau-like' scattering process. By this we mean that the photons are frequently absorbed and re-emitted by the electrically charged virtual particles, which are (on the average) accelerating towards the center of the large mass. When a virtual particle absorbs the real photon, a new photon is re-emitted after a small time delay in the same general direction as the original photon. This process is called photon scattering (figure 9). We will see that photon scattering is central to the understanding of space-time curvature.

The velocity of light in an ordinary moving medium is already known to differ from its value in an ordinary stationary medium. Fizeau (1851) demonstrated this experimentally with light propagating through a current of water flowing with a constant velocity. Later (1915), Lorentz identified the physics of this phenomena as being due to his microscopic electromagnetic theory of photon propagation. Einstein attributed this to the special relativistic velocity addition rule. In EMQG, we propose that in gravitational fields (and in accelerated motion) the moving water of Fizeau's experiment is now replaced by the accelerated virtual particles of the quantum vacuum. Like in the Fizeau experiment, photons scatter by the accelerated motion of the virtual particles of the quantum vacuum.

Imagine what would happen if Fizeau placed a clock inside his stream of moving water. Would the clock keep time properly, when compared to an observer with an identically constructed clock placed outside the moving water? Of course not! The very idea of this seems almost ridiculous. Yet we are expected to believe that the flow of virtual particles does not affect clocks and rulers under the influence of a gravitational field, as compared to the identical circumstance in far space. If Einstein knew the nature of the quantum vacuum at the time he proposed general relativity theory, he might have been aware of this connection between gravity, space-time curvature, and accelerated virtual particles.

In special relativity, we know the importance of the propagation of light in understanding the nature of space and time measurements. The definition of an inertial frame in space is a vast 3D grid of identically constructed clocks placed at regular intervals with a ruler. Therefore, we will closely examine the behavior of light near the earth.

In order to understand the connection between light propagation and space-time curvature near *a large gravitational field* with EMQG theory, we find it useful to review the behavior of a high-speed, non-relativistic test particle moving at 1/100 light velocity c near the earth. For the case where the test particle moves from the floor (distance r from the center of the earth) to the ceiling (height r+h, where h is small) on the *earth* (where the acceleration is 1g), we find that with an non-relativistic initial velocity $v_0 = 1/100$ c, the velocity of the particle at the detector is approximately ($v_0$ - gt). Here we can ignore all relativistic effects. The particle has a downward acceleration of 1g, which is *independent* of it's mass. Since $t \approx h/c$, we find that the final velocity of the particle with respect to the detector is given approximately by: **$v_0(1 - gh/c^2)$**, according to Newtonian physics. The reason for the change in the velocity of the high-speed particle (with respect to the detector) is that as the particle moves up, there is a change in the Newtonian gravitational potential on earth. This decelerates at –1g the particle as it moves towards the detector.

For the case where the same particle moves with velocity $v_0 \approx 1/100$ c from the floor of **an accelerated rocket** (same 1g) to the ceiling at the same small height h, the final velocity of the particle with respect to the detector is again ($v_0$ -gt), but for a *different* reason. As before, since $t \approx h/c$, and $v_0 = 1/100$ c, we find that the final velocity of the particle with respect to the detector is given approximately by: **$v_0(1 - gh/c^2)$** as in a gravitational field. The reason for the change in the velocity of the high speed particle (as viewed by the detector) is that as the particle moves up, it is now the detector itself which attains the velocity (-gt) with respect to the particle during the time t, due to the acceleration of the rocket. The end result is the same, but different physical processes are occurring.

When viewed from the principles of EMQG theory, there is really only **one** reason for the equality of the final rocket and earth velocities of our high-speed particle. In both cases, the elementary particles that make up the high-speed particle maintained a net statistical average *acceleration of zero* with respect to the virtual quantum vacuum particles. Thus, for the final velocity of the particle $v_0(1 - gh/c^2)$ on the rocket, the acceleration 1g represents the relative acceleration of the detector with respect to the high-speed particle which is in equilibrium with the non-accelerating quantum vacuum. And for the final velocity of the particle $v_0(1 - gh/c^2)$ on the earth, the acceleration 1g represents the relative acceleration of the detector with respect to the high speed particle which is also in equilibrium with the surrounding, but now falling, quantum vacuum virtual particles. Thus, on the earth, both the vacuum particles and the high-speed mass particle are *falling* at 1 g. Meanwhile inside the rocket it is only the detector that has the 1g acceleration with respect to the vacuum particles. We can see from this analysis that the equivalence of mass applies for the rocket and for the earth. We must stress that although equivalence exists, the physical process is actually different.

We will now take a bold step and assume that for the case on the surface of the earth the equation: $c(1 - gh/c^2)$ holds for the propagation of photons moving upwards, but only for *very short distances*. Technically this is true only at a point, which means that this equation must be written in differential form. We ignore the special relativistic postulate of the constancy of light velocity for now, and address this problem later. This means that photons continuously vary their velocity (the velocity of light is still an absolute constant between vacuum scattering events) by scattering with the falling virtual particles, as they propagate up or down. The scattering process will be described in detail later. If this picture is true, why is it that we do <u>not</u> observe this variation in light velocity in actual experiments on the earth?

First we must carefully understand what is meant by light velocity. Velocity is *defined* as distance divided by time, or $c=d/t$. Light has very few observable characteristics in this regard: we can measure velocity c (the ratio of d/t); frequency $\nu$; wavelength $\lambda$; and we can also measure velocity by the relationship $c=\nu\lambda$. It is important to note that all these observables are related. We know that $\nu = 1/t$ (t is the period of one light cycle) and $\lambda=d$ (the length of one light cycle). Thus, $c=d/t$ and $c=\nu\lambda$ are equivalent expressions. If we transmit green light to an observer on the ceiling of a room on the earth, and he claims that the light is red shifted, it is impossible for him to tell if the red shift was caused by the light velocity changing, or by space and time distortions which causes the timing and length of each of the light cycles to change. For example, if the frequency is halved, or $\nu_f = (1/2)\nu_i$ and the wavelength doubles $\lambda_f = 2\lambda_i$ (and you were not aware of both changes), then the velocity of light remains unchanged ($c=\nu\lambda$). However, if the velocity of light is halved, and you were not aware of it, then you could conclude that the frequency is halved, $\nu_f = (1/2)\nu_i$ and the wavelength doubles $\lambda_f = 2\lambda_i$. To illustrate this point, we will now examine what happens if an observer on the floor feeds a ladder (which represents the wave character of light) with equally spaced rungs to an observer on the ceiling, where each observer cannot see what the other observer does with the ladder.

Imagine a perfect ladder with equally space rungs of known length being passed up to you at a known velocity, such that it is impossible to tell the motion of the ladder other than by observing the rungs moving past you. If the rung spacing are made larger, you would conclude that either the ladder is slowing down, or that the spacing of the ladder rungs was increased. But it would be impossible to tell which is which. Let us assume that you make a measurement on the moving rungs, and observe a spacing of 1 meter between any two rungs. Then you observe that two rungs move past you every second. You therefore conclude the velocity of the ladder is 2 m/sec. Now, suppose that the ladder is fed to you at half speed or at 1 m/sec, and that you are not aware of this change in velocity. You could conclude that the velocity halved from your measurements, because you now observe that one rung appears in view for every second that elapses instead of two rungs, and that the velocity was thus reduced to 1 m/sec. However, you could just as well conclude that your space and time was altered, and that the velocity of the ladder is constant or unaffected. Since you observe only one rung in view per second instead of the usual two rungs, you could claim that the rung spacing on the ladder is enlarged (red-shifted) or doubled by someone, and that the velocity still remains unaltered. From this, you conclude that the frequency is halved, and that time measurements that will be based on this ladder are now dilated by a factor of two.

Which of these two approaches is truly correct? It is impossible to say by measurement, unless you know before hand what trait of the ladder was truly altered. For photons, the same problem exists. No known measurement of photons in an accelerated rocket or on the surface of the earth can reveal whether space and time is affected, or whether the velocity of light has changed. In EMQG theory, the variable light velocity approach is chosen for several reasons. First, the *equivalence of light motion in accelerated and gravitational frames now becomes **fully understood*** as a dynamic process having to do with motion (for gravity, hidden virtual particle motion), just as we found for ordinary matter in motion. Secondly, the *physical basis of the curvature* of Minkowski 4D space-time near a large mass now becomes clear. It arises from the interaction of light and matter with the background accelerated virtual particle processes. This process can be visualized as a fluid flow (for acceleration only) affecting the motion of light and matter. Finally, *the physical **action** that occurs between the earth and the surrounding space-time curvature now becomes clearly understood*. The earth acts on the virtual particles of the quantum vacuum through graviton exchanges, causing them to accelerate towards the earth. The accelerated virtual particles act on light and matter to produce curved 4D space-time effects. The physical process involved is photon scattering.

Since photon scattering is essential to our 4D space-time curvature approach we will examine scattering in some detail. First we review the conventional physics of light scattering in real moving, and real non-moving transparent matter such as water or glass. After this review, we will examine photon scattering due to the virtual particles of the quantum vacuum.

9.3     SCATTERING OF PHOTONS IN REAL, TRANSPARENT MATTER

It is a well known result of classical optics that light moves slower in glass than in air. Furthermore, the velocity of light in air is slower than that of its vacuum velocity. It also has been known for over a century that the velocity of light in a moving medium differs from its value in the same, stationary medium. Fizeau demonstrated this experimentally in 1851 (ref. 41). For example, with a current of water (with refractive index of the medium of n=4/3) flowing with a velocity V of about 5 m/sec, the relative variation in the light velocity is $10^{-8}$ (which he measured by use of interferometry). Fresnel first derived the formula (ref. 41) in 1810 with his ether dragging theory. The resulting formula relates the longitudinal light velocity '$v_c$' moving in the same direction as a transparent medium of an index of refraction 'n' defined such that 'c/n' is the light velocity in the stationary medium, which is moving with velocity 'V' (with respect to the laboratory frame), where c is the velocity of light in the vacuum:

Fresnel Formula: $v_c = c/n + (1 - 1/n^2) V$ (9.31)

Why does the velocity of light vary in a moving (and non-moving) transparent medium? According to the principles of special relativity, the velocity of light is a constant in the vacuum with respect to all inertial observers. When Einstein proposed this postulate, he was not aware of the fact that the vacuum is not empty. However, he was aware of Fresnel's formula and derived it by the special relativistic velocity addition formula for parallel velocities (to first order). According to special relativity, the velocity of light relative to the proper frame of the transparent medium depends only on the medium. The velocity of light in the stationary medium is defined as 'c/n'. Recall that velocities u and v add according to the formula: $(u + v) / (1 + uv/c^2)$
Therefore:

$v_c = [ c/n + V ] / [ 1 + (c/n) (V)/c^2 ] = (c/n + V) / ( 1 + V/(nc) ) \approx c/n + (1 - 1/n^2) V$ (9.32)

The special relativistic approach to deriving the Fresnel formula does not say much about the actual quantum processes going on at the atomic level. At this scale, there are several explanations for the detailed scattering process in conventional physics. Because light scattering is central to EMQG theory, we will investigate these different approaches in more detail below:

## 9.4 SCATTERING OF PHOTONS IN THE QUANTUM VACUUM

The above analysis can now be used to help us understand how photons travel through the virtual particles of the quantum vacuum. First we investigate the propagation of photons in the vacuum in far space, away from all gravitational fields. The virtual particles all have random velocities and move in random directions, and have random energies $\Delta E$ and life times $\Delta t$, which satisfies the uncertainty principle: $\Delta E \Delta t > h/(2\pi)$. Imagine a real photon propagating in a straight path through the virtual particles in a given direction. The real photon will encounter an equal number of virtual particles moving in a certain direction, as it does from the exact opposite direction. The end result is that the quantum vacuum particles do not contribute anything different than if all the virtual particles were at relative rest. Thus, we can consider the vacuum as some sort of stationary matter medium, with a very high density.

Is the progress of the real photon delayed as it travels through the quantum vacuum, where it encounters many electrically charged virtual particles? The answer to this question depends on whether there is a time delay between the absorption, and subsequent re-emission of the photon by a given virtual particle. Based on our arguments above, we postulate that the photon is delayed as it travels through the quantum vacuum (EMQG, Postulate #4, ref. 1). The uncertainty principle definitely places a lower limit on this time delay. In other words, according to the uncertainty principle the time delay cannot be exactly equal to zero! Our examination of the physics literature has not revealed any previous work on the time delay analysis of photon propagation through the quantum vacuum, or any evidence to contradict our hypothesis of photon vacuum delay (presumably because of the precedent set by Einstein's postulate of light speed constancy).

We will take the position that the delays due to photon scattering through the quantum vacuum reduces the 'raw light velocity $c_r$' (defined as the photon velocity between vacuum particle scattering) to the average light velocity 'c' in the vacuum of 300,000 km/sec that we observe in actual experiments. Furthermore, we propose that the quantum vacuum introduces a vacuum index of refraction 'n' such that $c = c_r / n$. What is the raw light velocity? It is unknown at this time, but it must be significantly larger than 300,000 km/sec. The vacuum index of refraction 'n' must be very large because of the high density of virtual particles in the vacuum. What happens if the entire quantum vacuum is accelerated? How does the motion of a photon get affected? These questions turn out to have a deep connection to space-time curvature.

## 9.5      PHOTON SCATTERING IN THE ACCELERATED QUANTUM VACUUM

Anyone who believes in the existence of the virtual particles of the quantum vacuum (which carry mass), will acknowledge the existence of an accelerated state of virtual particles of the quantum vacuum near any large gravitational field. The graviton-masseon postulate states that gravitons from the real masseons on the earth exchange gravitons with the virtual masseons (both the virtual masseons and anti-masseons), causing a downward acceleration. The virtual particles of the quantum vacuum (now accelerated by a large mass) acts on light (and matter) in a similar manner as a stream of moving water acts on light in the Fizeau effect. How does this work mathematically? Again, it is impossible to compute the interaction of an accelerated collection of virtual particles of the quantum vacuum with light exactly. However, a simplified model can yield useful results. We will proceed using the semi-classical model proposed by Lorentz, above. We have defined the raw light velocity '$c_r$' (EMQG, ref. 1) as the photon velocity in between virtual particle scattering. Recall that raw light velocity is the shifting of the photon information pattern by one cell at every clock cycle on the CA, so that in fundamental units it is an absolute constant. Again, we assume that the photon delay between absorption and subsequent re-emission by a virtual particle is '$\tau$', and the average distance between virtual particle scattering is 'l'. The scattered light velocity $v_c(t)$ is now a function of time, because we assume that it is constantly varying as it move downwards towards the surface in the same direction of the virtual particles. The virtual particles move according to: $a = gt$, where $g = GM/R^2$.

Therefore we can write the velocity of light after scattering with the accelerated quantum vacuum:

$$v_c(t) = c_r \ [1 + (gt\tau/l) \ (1 - gt/c_r)] / [1 + (c_r\tau/l) \ (1 - gt/c_r)] \tag{9.51}$$

If we set the acceleration to zero, or $gt = 0$, then $v_c(t) = c_r \ / \ (1 + c_r\tau/l) = c_r/n$. Therefore, $\tau/l = (n - 1)/c_r$. Inserting this in the above equation gives:

$$v_c(t) = [(c_r/n) + (1 - 1/n) gt (1 - gt/c_r)] / [1 - (1 - 1/n)(gt/c_r)] \approx c_r/n + (1 - 1/n^2) gt$$
to first order in $gt/c_r$. \hfill (9.52)

Since the average distance between virtual charged particles is very small, the photons (which are always created at velocity $c_r$) spend most of the time existing as some virtual charged particle undergoing downward acceleration. Because the electrically charged virtual particles of the quantum vacuum are falling in their brief existence, the photon *effectively* takes on the <u>same downward acceleration</u> as the virtual vacuum particles (postualte #4). In other words, because the index of refraction of the quantum vacuum 'n' is so large, and $c = c_r/n$ and we can write in equation 9.52:

$$v_c(t) = c_r/n + (1 - 1/n^2) gt = c + gt = c (1 + gt/c) \text{ if } n \gg 1. \tag{9.53}$$

Similarly, for photons going against the flow (upwards): $v_c(t) = c (1 - gt/c)$ (9.54)

We will see that this formula for the variation of light velocity near a large gravitational field leads to the correct amount of general relativistic space-time curvature (section 9.6).

Einstein, himself briefly considered the hypothesis of variable light velocity near gravitational fields shortly after releasing his paper on the deflection of light in gravitational fields (ref. 33. It would be interesting to contemplate what Einstein might have concluded if he new about the existence of virtual particles undergoing downward acceleration near a massive object (or in accelerated frames). Since Einstein was aware of the work by Fizeau on the effect of light velocity by a moving media, he might have been able to explain the origin of space-time curvature at the quantum level.

Now let us imagine that two clocks that are identically constructed, and each calibrated with a highly stable monochromatic light source in the same reference frame. These clocks keep time by using a high-speed electronic divider circuit that divides the light output frequency by "n" such that an output pulse is produced every second. For example, the light frequency used in the clock is precisely calibrated to $10^{15}$ Hz; this light frequency is converted in to an electronic pulse train of the same frequency, where it is divided by $10^{15}$ to give an electronic pulse every second. Another counter in this clock increments every time a pulse is sent, thus displaying the total time elapsed in seconds on the clock display. Now, let us place these two clocks in a gravitational field on earth with one of them on the surface, and the other at a height "h" above the surface. The clocks are compared every second to see if they are still running in unison by exchanging light signals. As time progresses, the clocks loose synchronism, and the lower clock appears to run slower. According to general relativity, light always maintains a constant speed, and space-time curvature is responsible for the difference in the timing of the two clocks. Recalling the accelerated Fizeau-like quantum vacuum fluid, we can derive the same time dilation effect by assuming that the light velocity has <u>exactly</u> the same downward acceleration component of the background falling quantum vacuum virtual particles.

9.6     SPACE-TIME CURVATURE FROM SCATTERING THEORY

We are now in a position to formulate the EMQG equations for the time dilation near a large gravitational mass based on the Fizeau-like quantum vacuum fluid. We assume that light is moving upward from the surface of the earth. As the photon moves upward from point r to point r+$\Delta$r it decelerates at -1g:

c( r+$\Delta$r ) = c( r )  (1 - g $\Delta$t  / c)                                             (9.61)

Since $\Delta$t  = $\Delta$r /c for small distances, we can now write:

c( r+$\Delta$r ) = c( r )  (1 - g $\Delta$r / $c^2$)                                          (9.62)

Since, g = GM/$r^2$ at point r above the center of the earth, we can write this as:

c( r+$\Delta$r ) = c( r )  (1 - GM $\Delta$r / $r^2$ $c^2$)                                 (9.63)

Since, the only observable property of light that we can be *sure* about is the red shift, and c =  ν λ, it follows:

ν( r+$\Delta$r ) = ν( r )  (1 - GM $\Delta$r / $r^2$ $c^2$)                                 (9.64)

from which the wavelength appears longer by the same factor, or

λ( r+$\Delta$r ) = λ( r )  (1 + GM $\Delta$r / $r^2$ $c^2$)                                 (9.65)

To find the total change in frequency from point r to infinity, we integrate GM $\Delta$r / $r^2$ $c^2$:

$\int_r^\infty$   GM /( $r^2 c^2$ ) dr   =   GM / ($r^2 c^2$ ) and therefore,                (9.66)

$$\nu(\infty) = \nu(r)\ (1 - GM/rc^2) \tag{9.67}$$

But, since $\nu = 1/t$ by definition, therefore time is affected as follows:

$$1/t(\infty) = (1/t(r))\ (1 - GM/(rc^2)) \tag{9.68}$$

Finally, we have:

$$t(r) = \left(1 - GM/(rc^2)\right) t(\infty) \tag{9.69}$$

which is the exactly the same expression for time dilation from the Schwarzchild metric.

Similarly, wavelength received at infinity is increased by the following expression:

$$\lambda(\infty) = \lambda(r)\ (1 + GM/rc^2) \tag{9.691}$$

Now, an observer at infinity can use the light signal from the surface of the earth to make measurements of distance in his reference frame at infinity. For example, suppose that in his own reference frame, a reference laser light source is used to measure a given reference length, and say that this corresponds to 1,000,000 wavelengths or $10^6\ \lambda_r$, where $\lambda_r$ is the reference wavelength. Subsequently, he uses the light received from the surface of the earth from an identically constructed reference laser light source ($\lambda_r$) to measure the same length, and finds that when he counts the standard 1,000,000 wavelengths the reference length has shortened (because of the wavelength increase). In general he concludes that the distances at infinity $s(\infty)$ are contracted by the amount:

$$s(\infty) = s(r)\ (1 - GM/rc^2) \tag{9.692}$$

compared to distances $s(r)$ on the surface of the earth. Finally, we can write:

$$s(r) = \left(1 - GM/(rc^2)\right)^{-1} s(\infty) \tag{9.693}$$

which is the exactly the same expression for length that we found from the Schwarzchild metric. This equation specifies the amount of distortion for rulers on the earth "$s(r)$" compared to rulers positioned at infinity "$s(\infty)$". Thus by postulating that it is the light velocity that is actually varying (and not space-time curvature), we are led to the same amount of red shift, and the same amount of space-time curvature.

We can now see that in order to formulate a theory of gravity involving observers with measuring instruments (such as clocks and rulers) we must take into account how these measurements are affected by the local conditions of the quantum vacuum. Our analysis above shows that quantum vacuum can be viewed as a Fizeau-like fluid undergoing downward acceleration near a massive object, which affects the velocity of light. Indeed, not only is the velocity of light affected, it is *all* the particle exchange processes including graviton exchanges. Therefore, we find that the accelerated Fizeau-like 'quantum vacuum fluid' effects all forces. This has consequences for the behavior of clocks, which are constructed with matter and forces. After all, nobody questions the fact that a mechanical clock submerged in moving water cannot keep proper time with respect to an external clock. Similarly, a clock near a gravitational field (with a Fizeau-like, quantum vacuum fluid flow inside the clock) also cannot be expected to keep proper time with respect to an observer outside the gravitational field. The accelerated Fizeau-like 'quantum vacuum fluid' moves along radius vectors directed towards the center of the earth, and thus has a specific direction of action. Therefore, the associated space-time effects should also work along the radius vectors (and not parallel to the earth). This is precisely the nature of curved 4D space-time near the earth.

For the case of light moving parallel to the earth's surface, the light path is the result of a tremendous number of photon to virtual particle scattering interactions (figure 9). Again in between virtual particle scattering, the light velocity is constant and 'straight'. The total path is curved as shown in figure 9. The path the light takes is called a geodesic in general relativity. In EMQG, this path simply represents the natural path that light takes through the accelerated vacuum. For the case of light moving parallel to the floor of the accelerated rocket (figure 10), the path for light is also the result of virtual particle scattering, but now the quantum vacuum is not in a state of relative acceleration. Therefore, the path is straight for the observer outside the rocket. The observer inside the rocket sees a curved path simply because he is accelerating upwards.

We now see why Einstein's gravitational theory takes the form that it does. Because of the continuously varying frequency and wavelength of the light with height, Einstein interpreted this as a variation of space and time with height. We postulated that the scattering of light with the falling vacuum changes the light velocity in absolute CA units, which cause the *measurements of space and time* to be affected. As we have already seen, these two alternative explanations ***cannot*** be distinguished by direct experimentation. This is why the principle of the constancy of light velocity is still a postulate in general relativity (through the acceptance of special relativity).

We are now in a position to understand the concept of the geodesic proposed by Einstein. ***The downward acceleration of the virtual electrically charged masseons of the quantum vacuum serves as an effective 'electromagnetic guide' for the motion of light (and for test masses) through space and time***. This 'electromagnetic guide' concept replaces the 4D space-time geodesics that guide matter in motion in relativity. For light, this guiding action is through the electromagnetic scattering process of section 9.5. For matter, the electrically charged virtual particles guide the particles of a mass by the electromagnetic force interaction that results from the relative acceleration. Because the quantum vacuum virtual particle density is quite high, but not infinite (at least about $10^{90}$ particles/m$^3$), the quantum vacuum acts as a very effective reservoir of energy to guide the motion of light or matter.

The *relative nature* of 4D space-time can now be easily seen. ***Whenever the background virtual particles of the quantum vacuum are in a state of relative acceleration with respect to an observer, the observer lives in curved 4D space-time.*** Why should the reader accept this new approach, when both approaches give the same result? The reason for accepting EMQG is that the action between a large mass and 4D space-time curvature becomes quite clear. The reason that 4D space-time is curved in an accelerated reference is also clear, and very much related to the gravitational case.

The relative nature of curved 4D space-time also becomes very obvious. An observer inside a gravitational field would normally live in a curved 4D space-time. If he decides to free-fall, he cancels his relative acceleration with respect to the quantum vacuum, and 4D space-time is restored to flat 4D space-time for the observer. The principle of general covariance no longer becomes a principle, but merely results for the deep connection between the quantum vacuum state for accelerated frames and gravitational frames. Last, but not least, the principle of equivalence is completely understood as a reversal of the (net statistical) relative acceleration vectors of the charged virtual masseons of the quantum vacuum, and real masseons that make up a test mass. We have seen EMQG at work for spherically symmetrical and non-rotating masses. What about the nature of the virtual particle acceleration field around an arbitrary mass distribution in any state of motion?

## 10. CONCLUSIONS

Based on a new quantum field theory of gravity theory called EMQG, we have illustrated the hidden quantum processes involved in Einstein's principle of equivalence. We found that almost the same processes also occurs in Newtonian Inertia. We found that gravity involves *two* pure force exchange processes. *Both* the photon and graviton exchanges occurring simultaneously inside a large gravitational field. Both particle exchange processes follow the particle exchange paradigm that was introduced in QED. We modified a new theory of inertia first introduced in ref. 5, which we call HRP inertia. In HRP inertia, inertia is the resulting electromagnetic force interaction of the charged 'parton' particles making up a mass with the background virtual photon field, which they called the zero point fluctuations (or ZPF). We modified HRP inertia, which we now call Quantum Inertia (or QI). This modification involves the introduction of a new particle of nature called the masseon, which always combines with other masseons to produce all the known fermion mass particles of the standard model. The masseon is electrically charged (as well as possessing mass-charge). Quantum Inertia is based on the idea that inertial force is due to the tiny electromagnetic force interactions (not fully defined at this time) originating from each charged masseon particle of real matter undergoing relative acceleration with respect to the virtual, electrically charged masseon particles of the quantum vacuum. These tiny forces are the source of the <u>total resistance force</u> to accelerated motion in Newton's law 'F = MA', where the sum of each of the tiny masseon forces equals the total inertial force. The exact detail of the tiny electromagnetic forces is not known, and hence this remains a postulate of EMQG.

We found that gravity also involves the same 'inertial' electromagnetic force component that exists in an accelerated mass, which reveals the deep connection between inertia and gravity. Inside large gravitational fields there exists a similar quantum vacuum process that occurs for inertia, where the roles of the real charged masseon particles of the mass and the virtual electrically charged masseons of the quantum vacuum are reversed. Now it is the charged virtual particles of the quantum vacuum that are accelerating, and the mass particles are at relative rest. Furthermore, the general relativistic Weak Equivalence Principle (WEP) results from this common physical process existing at the quantum level in both gravitational mass and inertial mass. Gravity involves *both* the electromagnetic force (photon exchanges) and the pure gravitational force (graviton exchanges) that are occurring simultaneously. However, for a gravitational test mass, the graviton exchange process (only found in minute amounts in inertial reference frames) occurring between a large mass, the test mass, and the surrounding vacuum particles upsets perfect equivalence of inertial and gravitational mass, with the gravitational mass being slightly larger. One of the consequences of this is that if a very large, and a tiny mass are dropped simultaneously on the earth, the larger mass would arrive slightly sooner. Since this is in violation of the WEP, the strong equivalence principle is no longer applicable.

Therefore, based on a new theory of quantum gravity called EMQG which eliminates the classical ideas of space-time and matter, we have discovered the hidden quantum interactions that occur in Newtonian inertia and the quantum machinery behind Einstein's Weak Equivalence Principle.

## 12. FIGURE CAPTIONS

The captions for the figures are shown below:

Figures 1 to 4: Schematic Diagram of the Principle of equivalence

Figure 5:    Virtual Particles Bonded under free fall in a Rocket (1g)

Figure 6:    Virtual Particles Bonded under free fall near the Earth

Figure 7:    Virtual Particle Pattern for the Earth and Moon in Free Fall near the Sun

Figure 8:    Virtual Particle Pattern for the Earth and Moon in Free Fall in a Rocket

Figure 9:    Motion of Real Photons in the Presence of Virtual Particles Near Earth

Figure 10:   Motion of Real Photons in Rocket Accelerating at 1g

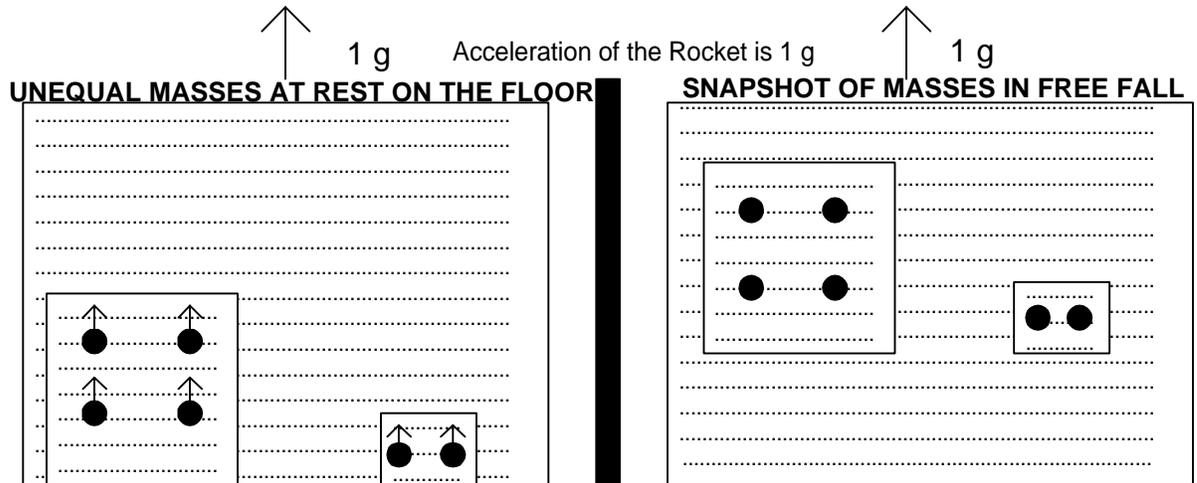

**Figure #1** - Masses '2M' and 'M' at rest on the floor of the rocket

**Figure #2** - Masses '2M' and 'M' in free fall inside of a rocket

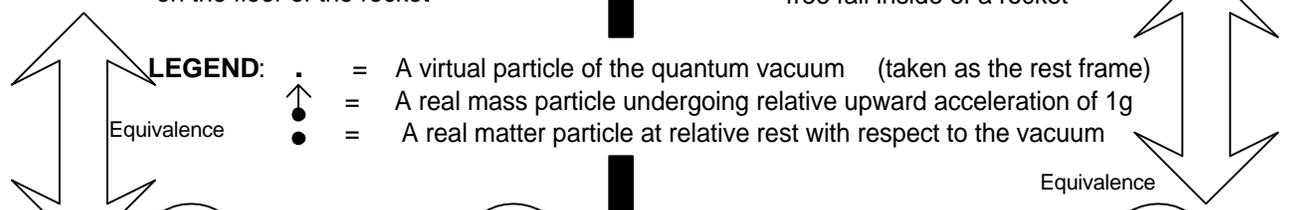

**LEGEND**:
- **.** = A virtual particle of the quantum vacuum (taken as the rest frame)
- ↑• = A real mass particle undergoing relative upward acceleration of 1g
- • = A real matter particle at relative rest with respect to the vacuum

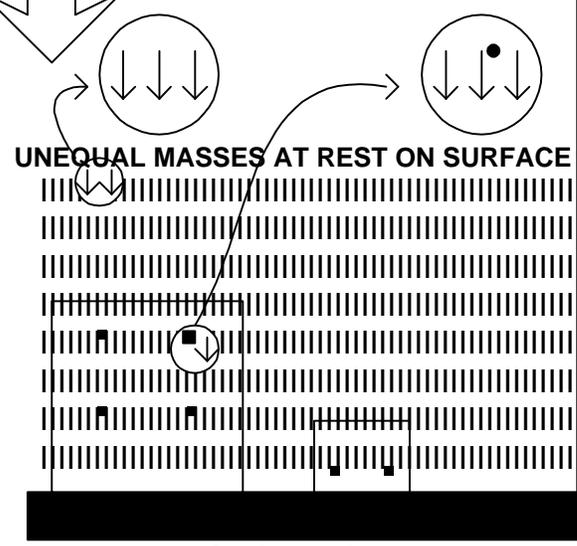
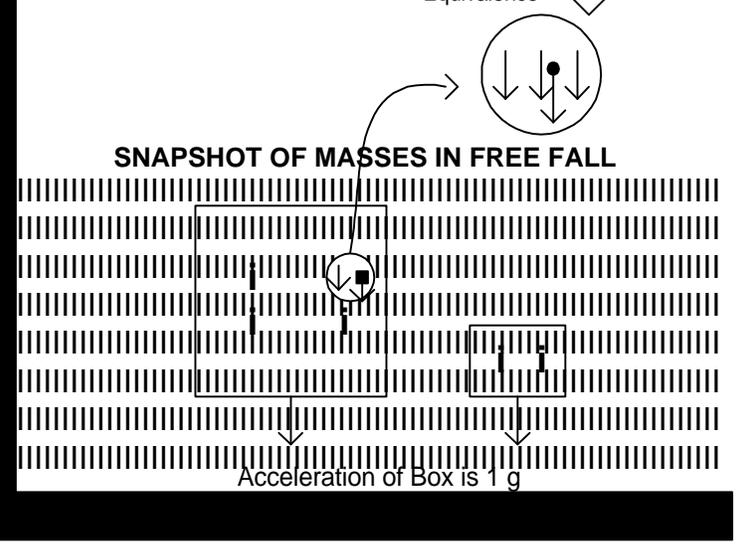

Surface of the Earth where gravity produces a 1 g acceleration

**Figure #3** - Masses '2M' and 'M' at rest on Earth's surface

**Figure #4** - Masses '2M' and 'M' in free fall above the Earth

**LEGEND:**
- **l** = Relative downward acceleration (1g) of a virtual particle
- **i** = Relative downward acceleration (1g) of a real matter particle
- **.** = A real stationary matter particle (with respect to the earth's center)

## SCHEMATIC DIAGRAM OF THE PRINCIPLE OF EQUIVALENCE

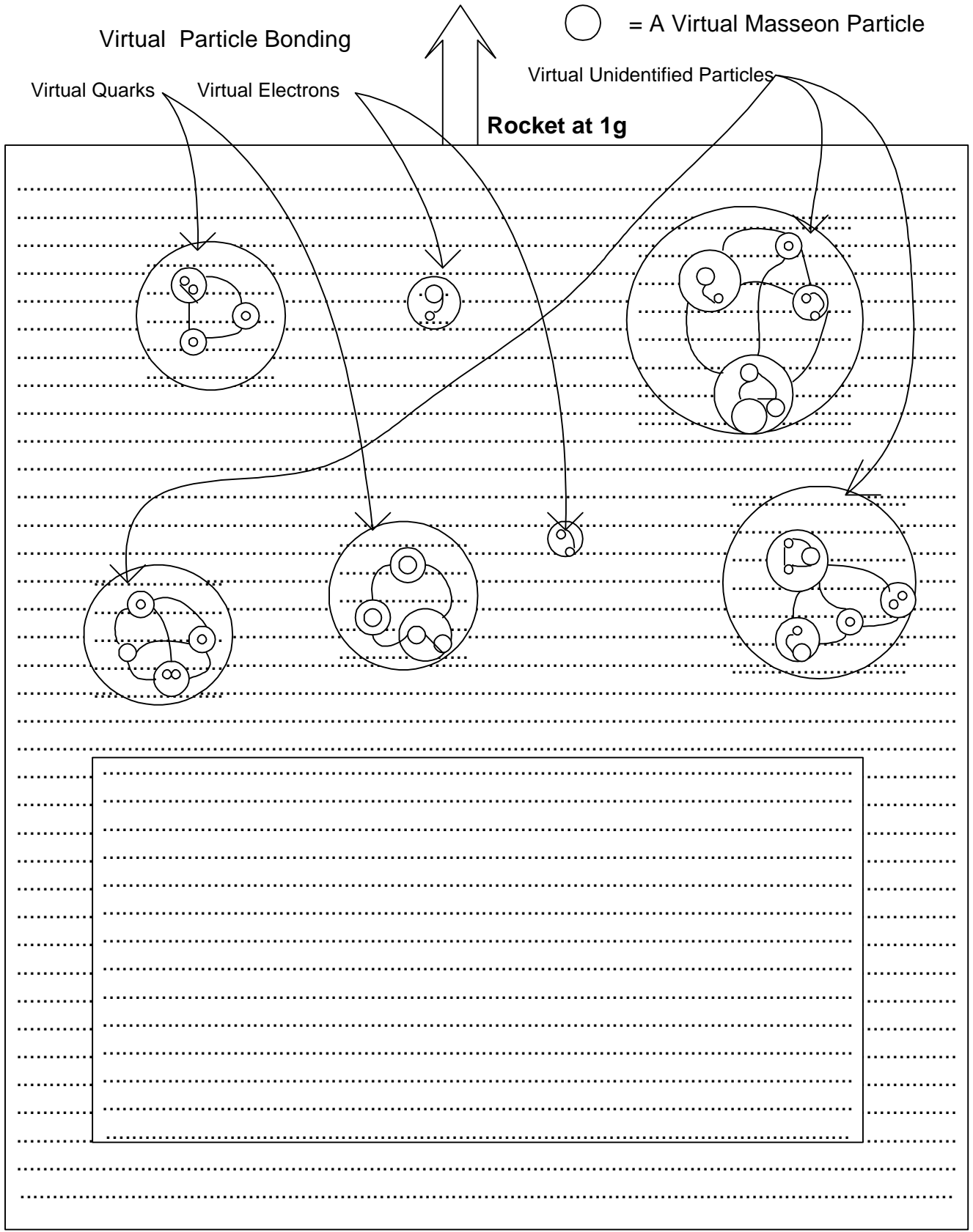

FIGURE #5 - VIRTUAL PARTICLES BONDED UNDER FREE FALL IN A ROCKET (1G)

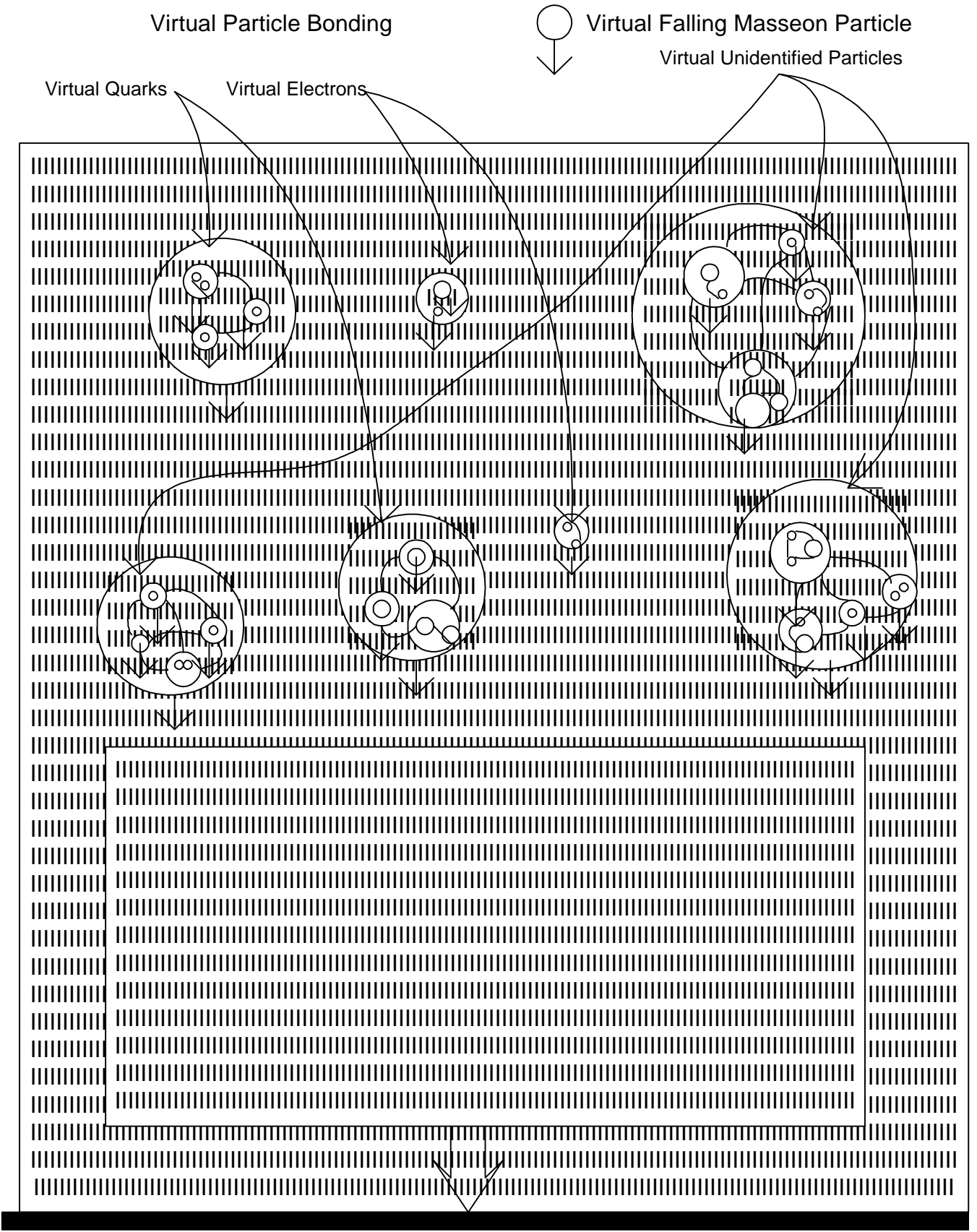

FIGURE #6 - VIRTUAL PARTICLES BONDED UNDER FREE FALL NEAR THE EARTH

## Figure #7 - VIRTUAL PARTICLE PATTERN FOR THE EARTH AND MOON IN FREE FALL NEAR THE SUN

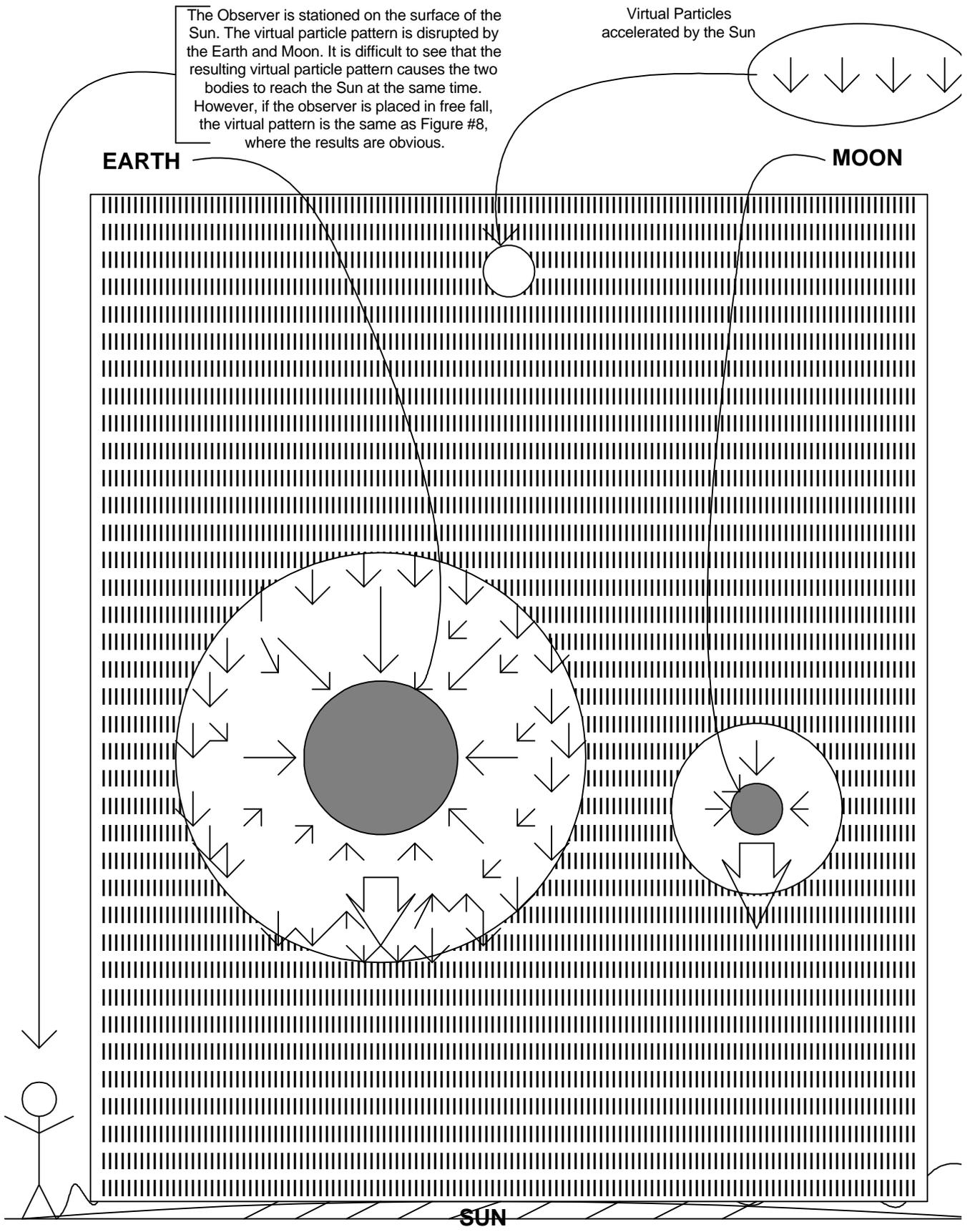

**Figure #8 - VIRTUAL PARTICLE PATTERN FOR THE EARTH AND MOON IN FREE FALL IN A ROCKET**

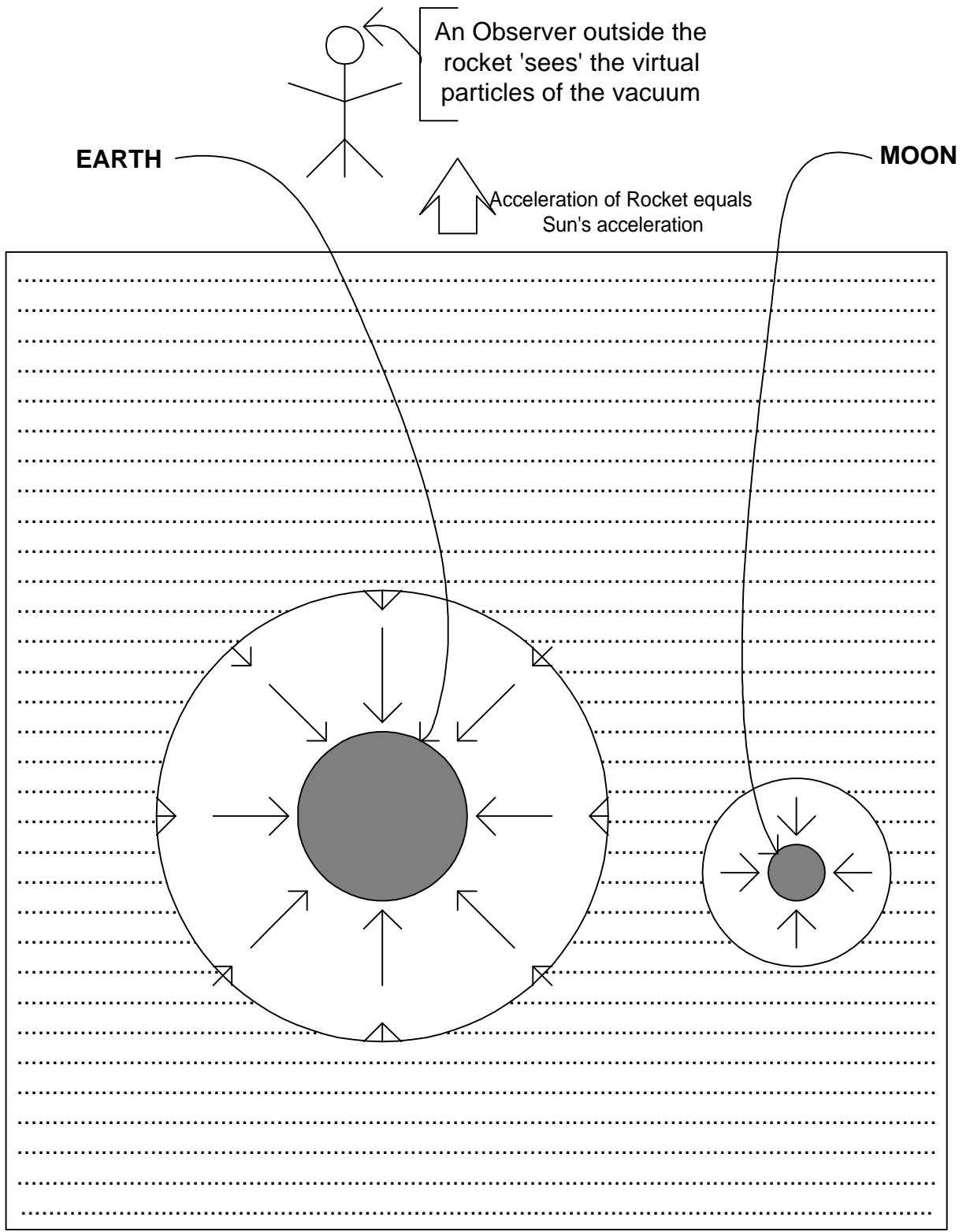

The Earth and Moon arrive on the floor of the huge rocket at the same time (the floor simply moves up to meet these bodies). But now, the virtual masseon particles near these two bodies are distorted and interacting with the real masseons in these two bodies.

## Figure #9 - MOTION OF REAL PHOTONS IN THE PRESENCE OF VIRTUAL PARTICLE NEAR EARTH

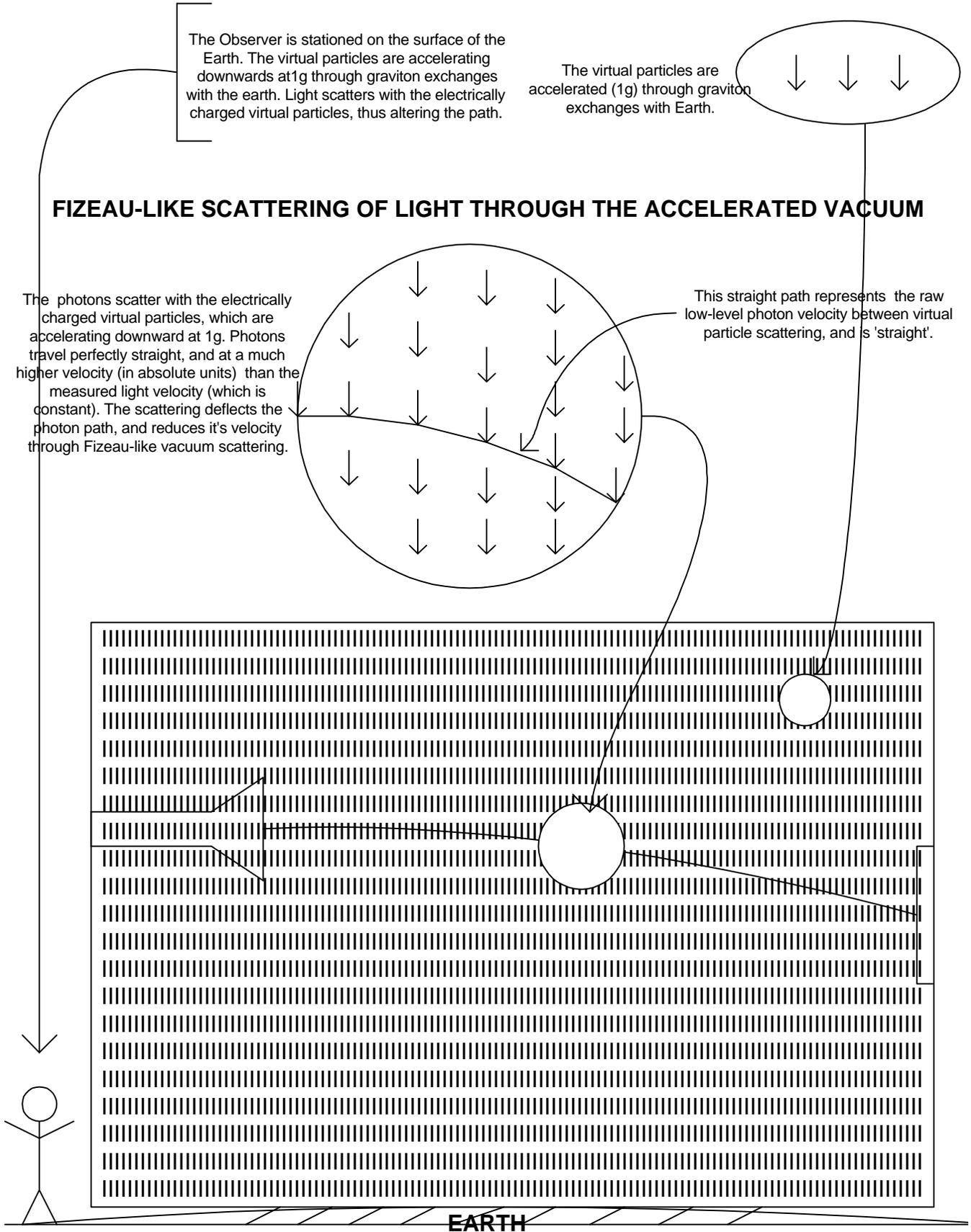

**Figure #10 - MOTION OF REAL PHOTONS IN A ROCKET ACCELERATING AT 1g**

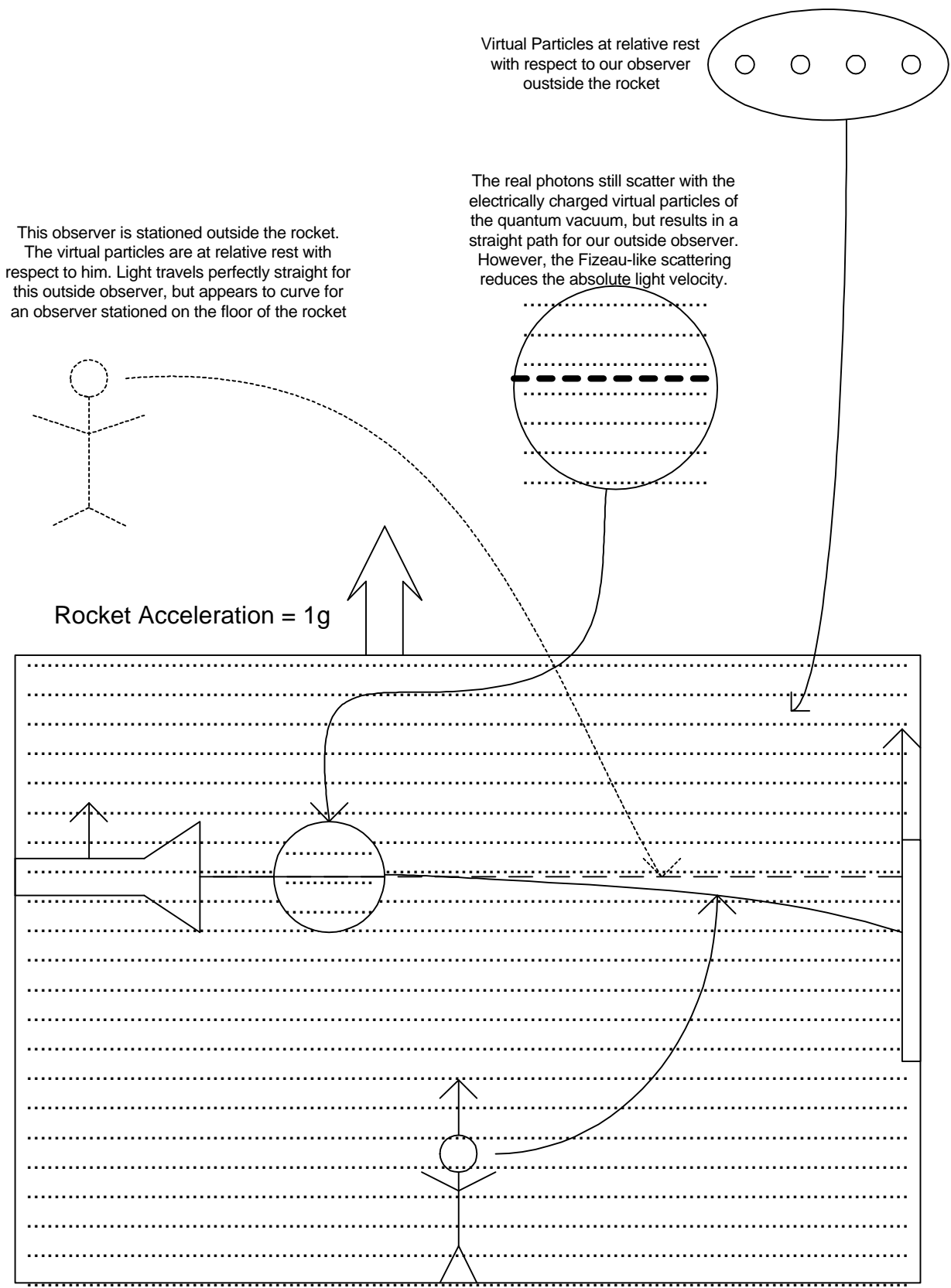